\documentclass[11pt,a4paper]{article}
\usepackage{jheppub}

\def\beq{\begin{equation}}
\def\eeq{\end{equation}}

\def\beq{\begin{equation}}
\def\eeq{\end{equation}}
\def\bea{\begin{eqnarray}}
\def\eea{\end{eqnarray}}

\def\eq#1{{Eq.~(\ref{#1})}}
\def\fig#1{{Fig.~\ref{#1}}}
\newcommand{\bas}{\bar{\alpha}_S}

\newcommand{\tr}{{\rm tr}}

\newcommand{\Lb}{\left(}
\newcommand{\Rb}{\right)}
\parskip=\itemsep               

\newcommand{\nn}{\nonumber}

\title{QCD unitarity constraints on Reggeon Field Theory.}
\author[a]{Alex Kovner,}
\author[b,c]{Eugene Levin,}
\author[d,a]{and Michael Lublinsky}
\affiliation[a]{Physics Department, University of Connecticut, 2152 Hillside Road, Storrs, CT 06269, USA}
\affiliation[b]{Departemento de F\'isica, Universidad T\'ecnica Federico Santa Mar\'ia, and Centro Cient\'ifico-\\
Tecnol\'ogico de Valpara\'iso, Avda. Espana 1680, Casilla 110-V, Valpara\'iso, Chile}
\affiliation[c]{Department of Particle Physics, Tel Aviv University, Tel Aviv 69978, Israel}
\affiliation[d]{Physics Department, Ben-Gurion University of the Negev, Beer Sheva 84105, Israel}

\abstract{We point out that the  $s$-channel  unitarity of  QCD imposes meaningful constraints on a possible form of the QCD Reggeon Field Theory. We show that neither the BFKL nor JIMWLK nor Braun's Hamiltonian satisfy the said constraints.  In a toy, zero transverse dimensional case we construct a model that satisfies the analogous constraint and show that at infinite energy it indeed tends to a "black disk limit" as opposed to the model with triple Pomeron vertex only, routinely used as a toy model in the literature.}

\begin{document}
\maketitle

\pagestyle{empty}
\newpage

\mbox{}

\pagestyle{plain}

\setcounter{page}{1}
\author{}

\abstract{ }
\dedicated{}
\preprint{}


\section{Introduction.}
Reggeon Field Theory (RFT) is an effective theory for description of hadronic scattering in QCD at asymptotically high energies. The basic ideas of RFT go back to Gribov \cite{gribov},
 and have been developed over the years in the context of QCD \cite{BFKL,glr,MUPA,MUDI,LIREV,LipatovFT, bartels,BKP, mv,Salam, KOLE,BRN,braun,BK}.   In its modern form, the QCD RFT in a certain limit has been identified \cite{reggeon}  with the so called JIMWLK evolution equation \cite{jimwlk}, or Color Glass Condensate (CGC)\cite{cgc}. The relevant limit is when a perturbative dilute projectile scatters on a dense target.
 
  Subsequently further relation between the CGC based approach and the RFT was explored. In particular recently we have shown that  one can generalize the JIMWLK Hamiltonian consistently in the regime where large Pomeron Loops are important \cite{ourbraun}. This regime includes the evolution of an initial dilute-dilute scattering to large rapidities, where at any given intermediate rapidity at most one of the evolved systems is dense. In this regime only Pomeron (Reggeon) splittings are important close to either one of the colliding objects, and one can write down a Hamiltonian, which encompasses both JIMWLK, and its dual (KLWMIJ\cite{klwmij}) evolution. The Hamiltonian in this regime contains the two triple Pomeron vertices, and (in the large $N_C$ limit) is the CGC equivalent of the Pomeron Lagrangian proposed by Braun \cite{braun} for description of the problem of scattering of large (but dilute) nuclei. So far, neither CGC nor RFT has been formulated in the most general case of scattering of two dense objects, although some work in this direction has been done\cite{KLduality,Balitsky05,SMITH,foam,GLV}.

 There are some significant differences between the original Gribov RFT framework and its QCD incarnation. The original reggeons in Gribov's RFT  are colorless, whereas the effective high energy degrees of freedom in QCD are frequently colored, such as reggeized  gluons \cite{gluon} or Wilson lines.  It  must be possible "to integrate over the color" and  reformulate QCD RFT in terms of  color neutral  exchange amplitudes, such as BFKL Pomeron \cite{BFKL}, however this has not yet been done explicitly.
QCD RFT in addition to the Pomeron contains  higher
order colorless Reggeons, such as quadrupoles and higher multipoles. Whether these higher Reggeons significantly affect high energy behavior of QCD amplitudes is not known at present. Finally, even if  the higher Reggeons can be discarded, it is not known whether the effective Pomeron Field Theory
has a finite number of transition vertices. The large $N_C$ limit of high energy QCD is a convenient setup for the study of these questions. In this paper we stick to the large $N_c$ limit and in fact restrict ourselves even further by considering the dipole model approach\cite{MUDI}, in which the Pomeron is the only relevant degree of freedom at high energy.

  It is clear by now that the CGC formalism conceptually provides a direct route to derive the Reggeon Field Theory from the underlying QCD. Due to this direct connection, one expects that it should be possible to understand some general features of RFT that are required by QCD.  
The current paper is devoted to discussion of how the unitarity of QCD as a fundamental field theory exhibits itself in the RFT framework.  To be precise we will be discussing the $s$-channel unitarity.
  

Our motivation to consider this question largely comes from earlier studies  of Pomeron Lagrangian proposed by Braun\cite{braun} for scattering of large 
(but dilute) nuclei. The Lagrangian incorporates the BFKL dynamics in the  linear regime  and contains two symmetric triple Pomeron vertices.
When the scattering amplitude is evolved within this framework to high enough rapidity, it exhibits paradoxical behavior: 
classical solutions to the  equations of motion bifurcate beyond some critical rapidity $Y_c$ \cite{motyka} and the dependence of 
the Pomeron amplitude on rapidity becomes unphysical. 
One is then left to wonder whether this peculiarity is a consequence of a possible non-unitarity of the Braun evolution.

The aim of this paper is to formulate the requirements of QCD  ($s$-channel) unitarity in the (Pomeron) RFT language.  
In short, the basic requirement of unitarity in RFT can be formulated as a certain property of the action of the RFT Hamiltonian on the projectile and target wave functions.

Both these wave functions are constructed as superpositions of (appropriate)
multi- dipole "Fock" states, the structure directly inherited from QCD.  The coefficients of the multi- dipole states, both in the projectile and target have the meaning of probabilities and hence each has to be positive and  smaller than one.  When acting on eitherthe  projectile or the target,  a unitary RFT Hamiltonian  has to preserve 
this property. This has to hold for all projectile/target states belonging to the corresponding Hilbert spaces.   

We will show that the above requirement of unitarity is not satisfied by the action of the  Braun Hamiltonian on either the projectile or the target wave function. The Balitsky-Kovchegov (BK) evolution \cite{BK} is partly unitary, in the sense that it unitarily evolves the projectile wave function. However its action on the target wave function strongly violates unitarity.  While we will not  discuss this in any detail, it is clear that the same conclusions hold  beyond the large $N_C$ approximation,  and 
thus both the JIMWLK\cite{jimwlk,cgc} and the KLWMIJ\cite{klwmij} Hamiltonians violate unitarity as well.  

Certain problems with  $t$-channel unitarity in the BK evolution have been already noticed a while ago in \cite{MShoshi}. Those were believed to have been cured by 
inclusion of Pomeron loops along the lines of Braun's construction \cite{braun,IAN,MUSH,LELU,SMITH, KLP,LMP,LEPP}.  Our present analysis shows that  problems with unitarity in the current RFT approaches run deeper. Although  the simple prescription {\it a la} Braun is likely sufficient to restore the $t$-channel untarity of the BK evolution, the $s$-channel unitarity is violated to some degree by all currrently available implementations of high energy evolution, including the Braun version of the BFKL 
Pomeron calculus.

We have made an attempt to find a modified RFT Hamiltonian which implements the unitarity conditions, and also reproduces the JIMWLK and KLWMIJ evolution in appropriate limits. This attempt was so far unsuccessful in the context of the realistic 2+1 dimensional RFT. An analogous program  however can be explicitly followed through in a toy model with zero transverse dimensions\cite{ACJ,AAJ,JEN,ABMC,CLR,CIAF,RS,KLremark2,SHXI,KOLEV,BIT,nestor,LEPRI}. 
The standard zero dimensional toy model with triple Pomeron vertex shares the paradoxical features of the Braun theory. In the context of a zero dimensional toy model we were able to construct a modified Lagrangian, which satisfies the zero dimensional analog of the QCD unitarity conditions. 
This model also has the JIMWLK (or rather its BK limit\cite{BK}) in the appropriate kinematics.We were also able to find explicit solutions for the evolution generated by this theory, and verify that it is free from paradoxes mentioned above.

The plan of this paper is the following. In Section 2 we recap the formulation of high energy evolution in the CGC approach. We also provide a path integral formulation of the calculation of the scattering amplitude, and demonstrate that in the appropriate limit it reproduces the Braun Lagrangian.
We also recap the peculiarities of the high energy evolution generated by this Lagrangian.

In Section 3 we confirm this strange behavior by considering small fluctuation analysis around  fixed points of the Braun Lagrangian.

In Section 4 we shift our attention to the zero dimensional model\cite{ACJ,AAJ,JEN,ABMC,CLR,CIAF,RS,SHXI,KOLEV,BIT,LEPRI} in order to demonstrate explicitly that it exhibits a similar paradoxical behavior. In this context we also demonstrate in a simple and straightforward way that the evolution of the zero dimensional analog of the Braun model as well as the JIMWLK model is not unitary.  Of course this statement has to be taken with a grain of salt. There is no fundamental field theory for which this model can serve  as an effective high energy limit. However the formal structure of the model is very similar to that of a realistic high energy QCD RFT. Thus in this context we can explore the formal analog of the QCD unitarity requirement in order to understand later its implementation in the realistic QCD RFT.

In Section 5 we construct a modification of the zero dimensional toy model which satisfies the unitarity requirements. We show that this model agrees with the JIMWLK and Braun Lagrangians in appropriate limits. We provide analytic solutions for the modified unitary model, and show that this evolution is devoid of the  worrisome features mentioned above.

In Section 6 we return to the 2+1 dimensional QCD RFT. We show that the Braun and JIMWLK evolutions are non-unitary in this realistic context. We also discuss difficulties we face in trying to follow through the program of constructing a unitary evolution in this dimensionality.

Finally Section 7  contains a short discussion of our results.

\section{Pomeron path integral from the CGC formalism.}
Our goal in this section is to derive a path integral representation for the scattering amplitude starting with the expressions derived in the CGC formalism in \cite{yinyang}. The main motivation for this reformulation is to make direct contact with the formulation of the RFT by Braun\cite{braun}.
\subsection{The scattering amplitude.}
In the CGC formalism the scattering of a fast moving projectile on a hadronic target is given by the expression
\beq
\label{scatm}
{\cal S}=\int d\rho d\alpha_T \delta(\rho)W_P[R]e^{i\int_z g^2\rho^a(z)\alpha_T^a(z)}\tilde W_T[\alpha_T]=\int d\rho \delta(\rho)W_P[R]W_T[S]
\eeq

Here $\rho^a(x)$ is the color charge density of the projectile, $\alpha^a_T(x)$ is the color field of the target, and
$R$ and $S$ are defined as
\beq \label{RS}
R_x=e^{t^a\frac{\delta}{\delta\rho^a_x}}; \ \ \ \ \ \ \ \ \ \ S_x=e^{ig^2t^a\alpha^a_x}
\eeq
with the projectile color field $\alpha^a(x)$ determined by the projectile color charge density $\rho^a(x)$ via solution of the static Yang-Mills equations.
The operator $R$ is the "dual Wilson line''. An insertion of a factor $R$ in the amplitude eq.(\ref{scatm}) is equivalent to appearance of an extra eikonal scattering factor associated with an additional parton. In this sense $R$ creates an additional parton in the projectile wave function. The Wilson line $S$ involves the projectile color field and has the meaning of the eikonal $s$-matrix of a {\it target} parton that scatters on the projectile.  
Here we have denoted the functional Fourier transform of $\tilde W_T[\alpha_T]$ by $W_T[S]$.

In this paper we will adhere to the dipole model framework\cite{MUDI}. We therefore assume, that all the observables can be written in terms of dipoles only, in which case, neglecting the possible contribution of Odderon,
the two basic elements of our calculation are the  Pomeron and its dual,
\beq \label{poms} P(x,y)= 1- \frac{1}{N_c}\tr[R_x R^{\dagger}_y]; \ \ \ \  \ \ \ \ \ \ \bar P(x,y)= 1- \frac{1}{N_c}\tr[S_x S^{\dagger}_y]\eeq
The integral over the charge density $\rho$ in eq.(\ref{scatm}) can be replaced by the integral over $\bar P$. In principle this change of variables involves a Jacobian, but it is inessential to our discussion and we will neglect it in the following.
Thus in the dipole model limit we have
\beq\label{scatmd}
{\cal S}=\int d\bar P \delta(\bar P)W_P[P]W_T[\bar P]
\eeq

The structure of the weight functions $W_P$ and $W_T$ is crucially important for the subsequent discussion of unitarity. This structure has been discussed in detail \cite{yinyang}. The presence of a physical dipole in the projectile wave function corresponds to a factor $d(x,y)\equiv 1-P(x,y)$ in $W_P$. Thus for a wave function that contains a distribution of dipole configurations (numbers and positions), the projectile weight function has the form
\beq\label{wp}
W_P=\sum_{n,\{x,\bar x\}}F^n(\{x,\bar x\})\prod_{i=1}^n[1-P(x_i,\bar x_i)]
\eeq
The functions $F^n(\{x,\bar x\})$ are probability densities, and therefore are nonnegative definite $F^n(\{x,\bar x\})\ge 0$.  Similarly, a dipole in the target wave function carries a factor 
$\bar d(x,y)\equiv 1-\bar P(x,y)$ in $W_T$, so
\beq\label{wt}
W_T=\sum_{n,\{x,\bar x\}}\bar F^n(\{x,\bar x\})\prod_{i=1}^n[1-\bar 
P(x_i,\bar x_i)]
\eeq
with
$\bar F^n(\{x,\bar x\})\ge 0$.
Furthermore, the weight functions $W_P$ and $W_T$ are normalized as:
\beq
\int d\bar P  \delta(\bar P) W_P[P]\,=\,1\,;\ \ \ \ \ \ \ \ \ W_T[0]\,=\,1
\eeq
which is equivalent to the proper normalization of the total probability
\beq
\sum_n\int_{\{x,\bar x\}}F^n(\{x,\bar x\})=1; \ \ \ \ \  \sum_n\int_{\{x,\bar x\}}\bar F^n(\{x,\bar x\})=1
\eeq
Considered as operators on the space of functionals $W$, the objects $P$ and $\bar P$ have nontrivial commutation relations. In principle those are directly calculable from the definitions eq.(\ref{RS}), but this is not a trivial calculation.  In the literature these commutation relations are usually approximated by those calculated in the dilute regime. In this regime, where any projectile dipole scatters only on a single target dipole (and vice versa), we can approximate $\bar P$ by \cite{reggeon}
\beq\label{pdagger}
P^\dagger (x,y)
=\frac{N_c^2}{4\pi^4 \bar{\alpha}_s^2}\nabla^2_x\nabla^2_y\bar P(x,y)
\eeq
where $\bar{\alpha}_s=\alpha_s N_C/\pi$ and
\beq\label{commut}
[P^\dagger(x,y),P(u,v)]=\delta^2(x-u)\delta^2(y-v)+\delta^2(x-v)\delta^2(y-u)
\eeq
Equivalently
\beq\label{barphi}
\bar P(x,y)\approx\bar\Phi(x,y)\equiv\int_{u,v}\gamma(x,y;u,v)P^\dagger(u,v)
\eeq
where $\gamma(x,y;u,v)$ is the Born level scattering amplitude  of a dipole $(x,y)$ on the dipole $(u,v)$
\beq 
\gamma(xy,uv)=\frac{\alpha_s^2}{8}\ln^2\frac{(x-u)^2(y-v)^2}{(x-v)^2(y-u)^2}
\eeq
The function $\gamma$ satisfies 
\beq\label{gamma2}
\nabla^2_x\nabla^2_y\gamma(xy,uv)=2\,\pi^2\,\alpha_s^2\,[\delta^2(x-u)\delta^2(y-v)+\delta^2(x-v)\delta^2(y-u)]
\eeq

With these commutation relations eqs.(\ref{commut},\ref{barphi}) the interpretation of the calculation of the scattering amplitude in eq.(\ref{scatmd}) is rather neat and intuitive.
Moving one operator $\bar P(x,y)$ from $W_T$ through $W_P$ one kills one of the operators $P(u,v)$ and instead acquires a factor $-\gamma(x,y;u,v)$, which is the dipole-dipole scattering amplitude. The original Pomeron $P(x,y)$ vanishes once it arrives next to the $\delta(\rho)$. The net result is that the Pomerons $P(x,y)$ and $\bar P(u,v)$ leave behind a factor $-\gamma(x,y;u,v)$ and disappear from the rest of the calculation, in accordance with the approximation that any dipole of the target can only scatter on one dipole of the projectile, and after doing so does not participate in any further scatterings. This clearly corresponds to dilute limit where a given projectile dipole can meet at most one target dipole while traversing the target.

In Sections 5 and 6 we will discuss the modification of the commutation relation between $P$ and $\bar P$ and an associated interpretation in terms of dipole-dipole scattering.

Eq.(\ref{scatmd}) defines the scattering matrix at some initial rapidity. The $S$-matrix evolved by the rapidity $Y$ is given by
\beq
\label{scatmde}
{\cal S}=\int d\bar P \delta(\bar P)\,W_P[P]\,e^{-H_{RFT}[P,\bar P]Y}\,W_T[\bar P]
\eeq
 Eq.(\ref{scatmde}), does not presuppose the commutation relation eq.(\ref{pdagger}) and remains valid if $\bar P$ is a more complicated function of the conjugate Pomeron $P^\dagger$, e.g. of the type we will discuss in subsequent sections.
The Hamiltonian $H_{RFT}$ is known in two limits - for dilute-dense\cite{klwmij} and dense-dilute\cite{jimwlk,cgc} situation. Another version of $H_{RFT}$ was suggested by Braun\cite{braun} as appropriate to description of scattering of two nuclei. We will discuss explicitly these Hamiltonians later, but for now our goal is to derive a general path integral representation for the $S$-matrix eq.(\ref{scatmde}).

\subsection{The path integral representation.}

First let us note that the expression eq.(\ref{scatmde}) is a multidimensional generalization of a ``quantum mechanical'' amplitude of the general form
\beq X= \int dx\delta(x)W_1(\hat p)e^{-\hat H(\hat p,x)Y}W_2(x)\eeq

Using the exponential representation of the $\delta$-function, and the fact that the only non-vanishing contributions come from terms where all derivatives in $W_1(\hat p)$ act on this exponential, we can write
\beq X=\int dx dp e^{ipx}W_1(p)e^{-\hat H(\hat p,x)Y}W_2(x)=\int dx dp W_1(p)W_2(x)\langle x\vert e^{-\hat H(\hat p,x)Y}\vert p\rangle\eeq

With the usual trick of inserting resolution of identity at intermediate ``times'' (a.k.a. ``rapidities), this can be written as the integral over trajectories with somewhat unusual boundary conditions
\beq X=\int dx dp W_1(p)W_2(x)\int_{x(Y)=x;\ \ p(0)=p} dx(\eta)dp(\eta)e^{-{S}};\ \ \ \ \ S=\int_0^Y d\eta [ip \frac{dx}{d\eta}-H]\eeq
Returning to the RFT, and using the correspondence
\beq x\rightarrow P(u,v); \ \ \ \ \ \ \ \hat p\rightarrow -iP^\dagger(u,v)\eeq
which follows from the commutation relation eq.(\ref{commut}), we can write\footnote{Strictly speaking 
$W_T[P^\dagger]\neq W_T[\bar P]$ and has to be renamed. Same remark applies to the Hamiltonian $H$. We will ignore these semantic differences.}
\beq\label{S}
{\cal S}=\int d P^\dagger dPW_P[P]W_T[P^\dagger]\int_{P(Y)=P;\ \ P^\dagger(0)=P^\dagger}DP(\eta)DP^\dagger(\eta) e^{\int^Y_0 d\eta \left[P^\dagger\frac{\partial P}{\partial\eta}-H(P,P^\dagger)\right]}
\eeq

 Throughout this paper, we will be focussing on several Pomeron Hamiltonians. The first one is the BK Hamiltonian, which is  a dipole/large $N_C$
version of the  KLWMIJ Hamiltonian:
\beq\label{HBK}
H_{BK}={\bar \alpha_s\over 2\pi}\,\int K(x,y|z)\,P^\dagger(x,y)\,\left[P(x,z)+P(z,y)-P(x,y)-P(x,z)P(z,y)\right]
\eeq
where $K(x,y|z)=\frac{(x-y)^2}{(x-z)^2(y-z)^2}$ is the BFKL kernel in the dipole form. The dual version of this Hamiltonian, which we will refer to  as the
$KB$ Hamiltonian, is a dipole/large $N_C$ version of the JIMWLK
\beq\label{HKB}
H_{KB}={\bar \alpha_s\over 2\pi}\,
\int K(x,y|z)\left[\bar P^\dagger(x,z)+\bar P^\dagger(z,y)-\bar P^\dagger(x,y)-\bar P^\dagger(x,z)\bar P^\dagger(z,y)\right]\,\bar P(x,y)
\eeq
Note that to write eqs.(\ref{HBK},\ref{HKB}) we did not have to assume a specific relation between $\bar P$ and $P^\dagger$ (or $P$ and $\bar P^\dagger$).

The third interesting Hamiltonian was written by Braun in \cite{braun}, and it explicitly assumes the dilute regime commutation relations eq.(\ref{pdagger}).

In order to write (\ref{S}) in the form presented in \cite{braun}, we introduce 
 a "linearized" version of the Pomeron by
\beq \label{dp}
d(x,y)=\exp[-\Phi(x,y)]; \ \ \ \ \ \ P(x,y)\approx \Phi(x,y)
\eeq
where the last approximate equality holds for small $P$, or dilute projectile limit. In the same approximation
\beq[\bar\Phi(x,y),\Phi(u,v)]=\gamma(x,y;u,v)
\eeq
We then have
\beq\label{action}
{\cal S}\simeq \int d \bar\Phi d\Phi W_P[\Phi]W_T[\bar\Phi]\int_{\Phi(Y)=\Phi;\ \ \bar \Phi(0)=\bar\Phi}D\Phi(\eta)D\bar \Phi(\eta) e^{\int^Y_0 d\eta \left[\frac{N_c^2}{4\pi^4\bar\alpha_s^2}[\nabla^2_x\nabla^2_y\bar\Phi(x,y)]\frac{\partial }{\partial\eta}\Phi(x,y)-H(\Phi ,\bar\Phi)\right]}
\eeq

As for the weight functions $W$, it follows from our previous discussion and in particular eqs.(\ref{wp},\ref{wt}) that they can be expressed in term of $\Phi$ and $\bar\Phi$. For a projectile and a target with fixed numbers of dipoles at given coordinates we have
\beq W_P=\exp[-\int J_P(x,y)\Phi(x,y)]; \ \ \ \ \ W_T=\exp[-\int J_T(x,y)\bar\Phi(x,y)]\eeq
where we have assumed that $P$ and $\bar P$ are both small at initial rapidity. The ``currents'' $J_P(x,y)$ and $J_T(x,y)$ are simply the number density of the dipole in the projectile and target at points $(x,y)$ respectively.

These weight functions can be traded for fixed boundary conditions on the Pomeron fields. Differentiating the action with respect to $\Phi$ gives the equation of motion for $\bar\Phi$ with the source term $J_P(x,y)\delta(\eta-Y)$. Similarly the equation of motion for $\Phi$ acquires the source term $J_T(x,y)\delta(\eta)$.
Integrating as usual the appropriate equation of motion across the appropriate boundary one finds that the presence  of the source terms is equivalent to imposing the boundary conditions
\beq
\Phi_{\eta=0}(x,y)=\phi(x,y)\equiv\int_{u,v}\gamma(x,y;u,v)J_T(u,v); \ \ \ \ \ \ \bar\Phi_{\eta=Y}(x,y)=\bar\phi(x,y)\equiv\int_{u,v}\gamma(x,y;u,v)J_P(u,v)
\eeq
These boundary conditions are equivalent to specifying the Born amplitude for the dipole scattering on the target and the projectile prior to rapidity evolution.

Eq.(\ref{action}) is the path integral representation considered by Braun \cite{braun} if the Hamiltonian is taken as
\begin{eqnarray}\label{braun}
H_B&=&\frac{N_c^2}{2\pi\bar\alpha_s}\int\bar \Phi(x,y)\nabla^2_x\nabla^2_y\left[K(x,y|z)[\Phi(x,z)+\Phi(z,y)-\Phi(x,y)-\Phi(x,z)\Phi(z,y)\right]\nonumber\\
&-&\Phi(x,y)\nabla^2_x\nabla^2_y\left[
K(x,y|z)\bar\Phi(x,z)\bar\Phi(z,y)\right]
\end{eqnarray}


\section{Peculiarities of the Braun evolution.}
In this section we discuss qualitatively the nature of solutions to the equations of motion generated by the Braun Hamiltonian. 
The classical equations of motion are given by
\bea
 \,\,\,\,\,& & \frac{\partial \Phi(x,y;\eta)}{\partial\,\eta}\,\,= \label{BAEQ1}\\
 &=&\,\,\frac{\bas}{2
\pi}\,\int\,d^2\,z\,K\left(x,y|z
\right)\,\Big\{ \Phi(x,z;\eta) \,+\,\Phi(z,y;\eta)\,-\,\Phi(x,y;\eta)\,- \Phi(z,y;\eta)\,\Phi(x,z;\eta) \Big\} \nonumber
\\
&-& \,\frac{\bas}{ 2\pi}\,\int_{z,x',y'}L^{-1}_{xy;x'y'} \,\,K \left(x',y'|z
\right)\,\,\left[\left\{L_{zy'} \,\Phi\left(z,y',\eta\right)\right\}\,\bar \Phi(x',z; \eta)+\left\{L_{zx'} \,\Phi\left(z,x',\eta\right)\right\}\,\bar \Phi(y',z; \eta)\right]\,\nonumber\\
\nn\\
\nn\\
 \,\,\,\,\,& & - \frac{\partial \bar \Phi(x,y;\eta)}{\partial\,\eta}\,\,= \label{BAEQ2}\\
 &=&\,\,\frac{\bas}{2
\pi}\,\int\,d^2\,z\,K\left(x,y|z
\right)\,\Big\{ \bar \Phi(x,z;\eta) \,+\,\bar \Phi(z,y;\eta)\,-\,\bar \Phi(x,y;\eta)\,- \bar \Phi(z,y;\eta)\,\bar \Phi(x,z;\eta) \Big\} \nonumber
\\
&-& \,\frac{\bas}{ 2\pi}\,\int_{z,x',y'}L^{-1}_{xy;x'y'} \,\,K \left(x',y'|z
\right)\,\,\left[\left\{L_{zy'} \,\bar \Phi\left(z,y',\eta\right)\right\}\, \Phi(x',z; \eta)+\left\{L_{zx'} \,\bar \Phi\left(z,x',\eta\right)\right\}\, \Phi(y',z; \eta)\right]\,\nonumber
\eea
The operator $L_{xy} \,=\,(x - y)^4\nabla^2_x \,\nabla^2_y$. 

These equations of motion are solved subject to boundary conditions $\Phi(\eta=0)=\phi; \ \ \bar \Phi(\eta=Y)=\bar \phi$, where $\phi$ and $\bar \phi$ are given finctions of the dipole sizes.
The equations have four fixed point $(\Phi,\bar \Phi)=(0,0),\ (0,1), \ (1,0)$ and $(1,1)$.

First let us clarify what we mean by the term ``fixed point''. The evolution equations are solved with boundary conditions on  $\phi(0)$ and $\bar \phi(Y)$. 
If the  values of  $\phi(0)$ and $\bar \phi(Y)$ are chosen to be exactly the fixed point values, the solution of the equations of motion is equal to these values for all $\eta$; $\Phi(\eta)=\phi(0); \ \ \bar \Phi(\eta)=\bar\phi(Y)$.
If the values of $\phi$ and $\bar \phi$ are not chosen to be exactly the ``fixed point'' values it is obviously impossible for the solution to reach any of the fixed points at all rapidities. However one might expect that if the evolution interval in rapidity is very large, $Y\rightarrow\infty$, the solution will be arbitrarily close to one of the fixed point values  for large interval of intermediate rapidities $\eta$ of the length of order $Y$ away from the end points. This expectation may be too naive, and in fact we will see that in the zero dimensional toy model it is not strictly satisfied. However one certainly does expect that starting from the end points and moving towards the midpoint of the rapidity interval, the solution will develop towards one of the attractive fixed points, even though it may not quite reach it.

 Since the interpretation of $\Phi$ and $\bar \Phi$ is that of the scattering amplitude of an external dipole on the target and the projectile respectively, the point $(0,0)$ is the vacuum fixed point, where the scattering amplitude on both the target and the projectile vanishes. It is a repulsive fixed point, meaning that the solution of equations of motion departs from it with rapidity, given an initial condition which is not exactly zero. 
 
 The point $(1,1)$ corresponds to the dense-dense limit, where both the target and the projectile are black. Since one expects both the target and the projectile states to become dense as a result of the evolution, one expects this to be an attractive fixed point. In other words we expect that for boundary conditions $\phi\ne 0$ and $\bar\phi\ne 0$ and for a very large rapidity interval $Y$, the solution will be approaching  the point $\Phi(\eta)=1$ and $\bar \Phi(\eta)=1$ towards the middle of the rapidity interval  $0<\eta<Y$.  
 
 Finally we will refer to the points $(0,1)$ and $(1,0)$ as the ``BK'' fixed points, since they correspond to the situation where one of the colliding objects is dense and one is dilute. Naively we expect these two fixed points to be repulsive, albeit not as strongly repulsive as $(0,0)$. In other words, if the boundary conditions are not too far from these values, e.g. $\Phi(0)=\epsilon$, $\bar \Phi(Y)=1-\epsilon'$, for an intermediate range of $Y$ the solution will be close to the point $(0,1)$ for most intermediate rapidities $\eta$. However if $Y$ is increased to sufficiently large value, the solution for these boundary conditions will eventually flow towards $(1,1)$ at intermediate values of $Y\gg \eta\gg 0$.  

Our goal in this section is to determine whether our intuitive expectation on the nature of the fixed points is born out by the equations of motion of the Braun model.
In the following we concentrate on the points $A=(1,1)$ and $B=(1,0)$.
\subsection{Point A.}
First, let us take $\phi$ and $\bar \phi$ both close to unity, which means that already at initial rapidity both the projectile and the target are dense objects (nuclei). It is then reasonable to expect that $\Phi$ and $\bar \Phi$ stay close to unity in the whole rapidity interval $0<\eta<Y$. 

 We can then write down the linear equation for the deviation of $\Phi$ and $\bar \Phi$ from unity, $\Delta\equiv 1-\Phi$, $\bar \Delta \equiv 1-\bar \Phi$.
\bea
 \,\,\,\,\,& & \frac{\partial \Delta(x,y;\eta)}{\partial\,\eta}\,\,= \label{AEQ1}\\
 &=&\,\,\frac{\bas}{2
\pi}\,\int_z\,\, K\left(x,y|z
\right)\,\Big\{ \Delta(x,z;\eta) \,+\,\Delta(z,y;\eta)\,-\,\Delta(x,y;\eta)\,- \Delta(z,y;\eta)\, -\Delta(x,z;\eta) \Big\} \nonumber
\\
&-& \,\frac{\bas}{ 2\pi}\int_{zx'y'}\,L^{-1}_{xy;x'y'} \,\,K \left(x',y'|z
\right)\,\,\left\{L_{zy'} \,\Delta\left(z,y',\eta\right)+L_{zx'} \,\Delta\left(z,x',\eta\right)\right\}\,\,\nonumber\\
 \,\,\,\,\,& & - \frac{\partial \bar\Delta(x,y;\eta)}{\partial\,\eta}\,\,= \label{AEQ2}\\
 &=&\,\,\frac{\bas}{2
\pi}\,\int_z\,K\left(x,y|z
\right)\,\Big\{ \bar\Delta(x,z;\eta) \,+\,\bar\Delta(z,y;\eta)\,-\,\bar\Delta(x,y;\eta)\,- \bar\Delta(z,y;\eta)-\,\bar\Delta(x,z;\eta) \Big\} \nonumber \\
&-& \,\frac{\bas}{ 2\pi}\int_z\,L^{-1}_{xy;x'y'}\,K \left(x',y'|z
\right)\,\,\left\{L_{zy'} \,\bar\Delta\left(z,y',\eta\right)+L_{zx'} \,\bar\Delta\left(z,x',\eta\right)\right\}\nonumber
\end{eqnarray}
or after obvious cancellations
\bea
\,\,\,\,\,& & \frac{\partial \Delta(x,y;\eta)}{\partial\,\eta}\,\,= \label{1AEQ1}\\
 &=&\,\,-\frac{\bas}{2
\pi}\,\Big\{\int_z\,\,K\left(x,y|z\right)\,\,\Delta(x,y;\eta)\,+\int_{zx'\eta}\,L^{-1}_{xy;x'y'} \,\,K \left(x',y'|z
\right)\,\,\left\{L_{zy'} \,\Delta\left(z,y',\eta\right)+L_{zx'} \,\Delta\left(z,x',\eta\right)\right\} \Big\} \nonumber
\\
\,\,\,\,\,& & - \frac{\partial \bar\Delta(x,y;\eta)}{\partial\,\eta}\,\,= \label{1AEQ2}\\
&=&\,\,-\frac{\bas}{2
\pi}\,\Big\{\int_z\,K\left(x,y|z\right)
\,\,\bar\Delta(x,y;\eta)\,+\int_z\,L^{-1}_{xy;x'y'}\,K \left(x',y'|z
\right)\,\,\left\{L_{zy'} \,\bar\Delta\left(z,y',\eta\right)+L_{zx'} \,\bar\Delta\left(z,x',\eta\right)\right\}\Big\}\, \nonumber
\eea
It is more transparent to multiply these equations by the operator $L$ and write them as equations for $n(xy)=L(xy)\Delta(xy)$ and $\bar n(xy)=L(xy)\bar \Delta(xy)$. 
 The physical meaning of $n$ (similarly $\bar n$) is that of the logarithmic dipole density.
Using the fact that the BFKL kernel commutes with the operator $L$  we obtain:
\bea
\,\,\,\,\,& & \frac{\partial n(x,y;\eta)}{\partial\,\eta}\,\,= \label{1AEQ12}\\
 &=&\,\,-\frac{\bas}{2
\pi}\,\Big\{\int_z\,\,K\left(x,y|z\right)\,\,n(x,y;\eta)\,+\int_{z}\, \,\,K \left(x,y|z
\right)\,\,\left\{n\left(z,y,\eta\right)+n\left(z,x,\eta\right)\right\} \Big\} \nonumber
\\
\,\,\,\,\,& & - \frac{\partial \bar n(x,y;\eta)}{\partial\,\eta}\,\,= \label{1AEQ22}\\
&=&\,\,-\frac{\bas}{2
\pi}\,\Big\{\int_z\,K\left(x,y|z\right)
\,\,\bar n(x,y;\eta)\,+\int_z\,\,K \left(x,y|z
\right)\,\,\left\{\bar n\left(z,y,\eta\right)+ \,\bar n\left(z,x,\eta\right)\right\}\Big\}\, \nonumber
\eea
All the kernels on the RHS of these equations are positive, and thus we conclude that $\frac{\partial} {\partial\,\eta}n(x,y;\eta)<0$ and $\frac{\partial }{\partial\,\eta}\bar n(x,y;\eta)>0$. This means that $\Phi$ approaches unity as the rapidity $\eta$ increases, while $\bar \Phi$ approaches unity as $\eta$ decreases. Thus the fixed point $A$ is attractive, namely if we start with initial conditions where $\Phi$ is close to unity at $\eta=0$ and $\bar \Phi$ is close to unity at $\eta=Y$, and $Y$ is large enough, then at all values of $\eta$ the solution will be close to the fixed point. This is in accordance with our naive expectation.

\subsection{Point B.}
Now consider point $B$. 
Assume that our initial conditions fix $\Phi$ to be close to one at $\eta=0$, but fix $\bar \Phi$ to be small at $\eta=Y$.
Denoting $1-\Phi\equiv \Delta,\ \ \ \bar \Phi\equiv \bar\Delta$, we have the small fluctuation equations as
\bea
 \,\,\,\,\,& & \frac{\partial \Delta(x,y;\eta)}{\partial\,\eta}=-\,\,\frac{\bas}{2
\pi}\,\int_z\,\, K\left(x,y|z
\right)\,\,\Delta(x,y;\eta) \label{2AEQ1}
\\
 \,\,\,\,\,& & - \frac{\partial \bar\Delta(x,y;\eta)}{\partial\,\eta}\,\,= \label{2AEQ2}\\
 &=&\,\,\frac{\bas}{2
\pi}\,\int_z\,K\left(x,y|z
\right)\,\Big\{ \bar\Delta(x,z;\eta) \,+\,\bar\Delta(z,y;\eta)\,-\,\bar\Delta(x,y;\eta)\, \Big\} \nonumber \\
&-& \,\frac{\bas}{ 2\pi}\int_z\,L^{-1}_{xy;x'\eta}\,K \left(x',\eta|z
\right)\,\,\left\{L_{z\eta} \,\bar\Delta\left(z,\eta,\eta\right)+L_{zx'} \,\bar\Delta\left(z,x',\eta\right)\right\}\nonumber
\end{eqnarray}

The first equation, as before says that $\Phi$ approaches unity at $Y>0$. 

The second equation when rewritten in terms of $\bar n$ reads
\begin{equation}\label{tilden}
- \frac{\partial \bar n(x,y;\eta)}{\partial\,\eta}\,\,= -\frac{\bas}{2
\pi}\,\int_z\,K\left(x,y|z
\right)\,\,\bar n(x,y;\eta)\,
\end{equation}
This shows that $\bar n$ increases towards positive rapidities. If it starts off as small at $\eta=Y$ it becomes even smaller at $\eta<Y$ and thus this fixed point also is attractive.

This simplified analysis suggests that both A and B are attractive fixed points, and thus depending on the initial conditions, the system flows to one or the other at intermediate rapidities. 
This result is surprising and goes against our intuition. It suggests that for a physical situation where the dense-dilute scattering process is evolved to large rapidities, the dilute object never gets dense. 

This  interpretation of the behavior we have just found is not quite adequate, as we will discuss in the next sections. Nevertheless this behavior is counter intuitive. 
However our analysis is incomplete, since we have been a little cavalier about the dependence of the Pomerons on transverse coordinates. Although in the strict sense the points $A$ and $B$ are indeed all the fixed points of the flow, one never starts with initial condition which is close to saturation at all values of dipole size. A more careful analysis should allow for existence of a finite saturation momentum. It is logically possible that accounting for finite $Q_s$ will change the attractive nature of the point B. We will now perform this analysis.

\subsection{Point B - finite $Q_s$.}
Let us consider again vicinity of the point $B$. Let us assume that at any rapidity $\eta$, $\Phi(x-y)=1$ for $|x-y|>Q_s^{-1}(\eta)$, but $\Phi(x-y)$ is small for $|x-y|<Q_s^{-1}(\eta)$. We will still assume that $\bar \Phi$ is small for all dipole sizes. The value of the saturation scale $Q_s$  depends on $\eta$.

Consider  eq.(\ref{BAEQ1}) for the Pomeron $\Phi$.  
Here the contribution of the second line in eq.(\ref{BAEQ1}) is always second order in smallness, since $\bar \Phi$ is small and $L$ annihilates the constant part of $\Phi$.
Thus the equation of motion in this approximation becomes the BK equation whose behavior is well understood.

Consider first small external dipole sizes $x-y<Q_s^{-1}(\eta)$.
The contribution of the nonlinear term from large emitted dipole sizes $z>Q_s^{-1}$ cancels half of the contribution of the real part of the BFKL kernel, while the region of small $z$ leaves the contribution of full BFKL equation
\begin{eqnarray}\label{abc}
\frac{\partial \Phi(x,y;\eta)}{\partial\,\eta}|_{|x-y|<Q^{-1}_s(\eta)}\,\,&=&\frac{\bas}{2
\pi}\,\int_{|z|<Q_s^{-1}}\,K\left(x,y|z
\right)\,\Big\{ \Phi(x,z;\eta) \,+\,\Phi(z,y;\eta)\,-\,\Phi(x,y;\eta)\, \Big\}\\
&+&\frac{\bas}{2
\pi}\int_{|z|>Q_s^{-1}}\,K\left(x,y|z\right)[1-\Phi(x,y;\eta)]\nonumber
\end{eqnarray}
If we were to neglect the last term, the solution would be just that of the BFKL equation, namely exponentially growing with rapidity. The last term is also positive, and thus speeds up the evolution slightly. This term however is only important when the Pomeron is in the color transparency regime.
Recall that the BFKL kernel decreases as $z^{-4}$ at large $z$. Thus the contribution of the last term is proportional to $\alpha_s|x-y|^2Q_s^{2}=\alpha_s^3|x-y|^2\mu^2$, where $\mu^2$ is the gluon density.
The first term is the BFKL equation with gluon emissions limited to the short distance.
Its contribution can be estimated approximately as 
\beq \int_{|z|<Q_s^{-1}}\,K\left(x,y|z
\right)\,\Big\{ \Phi(x,z;\eta) \,+\,\Phi(z,y;\eta)\,-\,\Phi(x,y;\eta)\, \Big\}\sim \omega \Phi(x,y;\eta)\eeq
with $\omega$ a number of order unity. Eq.(\ref{abc}) then effectively reads
\beq \frac{\partial \Phi(x,y;\eta)}{\partial\,\eta}\sim\frac{\bas}{2
\pi}\,[\omega \Phi(x,y;\eta)+\alpha_s^2|x-y|^2\mu^2]\eeq
Thus once the Pomeron surpasses its color transparency limit, the second term is negligible and the evolution is dominated by the first term which is just the BFKL evolution.

For large dipoles, $|x-y|> Q_s^{-1}(\eta)$ the nonlinear term in eq.(\ref{BAEQ1}) cancels the contribution form the real part of BFKL kernel in the region $|z-x|>Q^{-1}_s(\eta)$ and $|z-y|>Q^{-1}_s(\eta)$.
In this large $z$ region we can write as before $\Phi(x,z)=1-\Delta(x,z)$ etc. The only contribution to the RHS of eq.(\ref{BAEQ1}) in this region is then
\beq \frac{\partial \Delta(x,y;\eta)}{\partial\,\eta}|_{|x-y|>Q^{-1}_s(\eta)} \,\,= -\frac{\bas}{2
\pi}\,\int_{ \{|z-x|>Q_s^{-1}(\eta); |z-y|>Q_S^{-1}(\eta)\}}\,d^2\,z\,K\left(x,y|z
\right)\,\,\Delta(x,y;\eta)\eeq

The ``small $z$'' region is split in two: $|z-x|<Q_s^{-1}(\eta)$ and $|z-y|<Q_S^{-1}(\eta)$. In the first one $\Phi(z-x)$ is small, but we can write $\Phi(z-y)=\Phi(x-y)=1-\Delta(x-y)$. Substituting this into the RHS of  eq.(\ref{BAEQ1}) we find complete cancellation to linear order in $\Delta$. The same happens in the second region of small $z$.   The only leftover is the virtual term integrated over the remainder of the space $|z-x|>Q^{-1}_s(\eta)$, $|z-y|>Q^{-1}_s(\eta)$. Since the integral of the virtual term is cut off in the infrared in the range $|z-x|\sim |x-y|$, the equation for small fluctuations in the saturation region becomes
\begin{equation}
\frac{\partial \Delta(x,y;\eta)}{\partial\,\eta} |_{|x-y|>Q^{-1}_s(\eta)} \,\,=-\frac{\bas}{
\pi}\,\ln [(x-y)^2Q_s^2(\eta)]
\Delta(x,y;\eta)\, 
\end{equation}
This is of course, the standard Levin-Tuchin argument \cite{LETU}. It shows that as the rapidity increases, the Pomeron approaches saturation. At small values it grows toward saturation according to BFKL, while close to saturation, it continues to grow albeit slowly. Since $Q_s(\eta)$ grows with $\eta$, more and more dipole sizes are saturated as rapidity increases.

The more interesting and problematic equation is the one for $\bar \Delta\equiv \bar \Phi$ (\ref{BAEQ2}). Our main interest here is to see whether allowing for a finite saturation momentum can reverse the flow of $\bar \Phi$ and somehow through a back door cause it to grow towards the smaller values of rapidity $\eta<Y$.

We consider the initial condition where $\bar\Phi$ is small at $\eta=Y$. It's saturation momentum in this rapidity range is vanishing.
 However in the evolution  we have to account for the effect of the finite saturation momentum of $\Phi$. 
 Taking this into account we find that the contribution to the last two terms in eq.(\ref{2AEQ2}) is restricted to the integration region $|z-x'|>Q_s^{-1}(\eta)$ and $|z-\eta|>Q_s^{-1}(\eta)$ respectively, since in the rest of the domain this contribution is quadratic in $\Delta \bar \Delta$.
 
Consider first large external dipoles $|x-y|> Q_s^{-1}$. In this regime (commuting as before the operator $L^{-1}$ with the BFKL kernel), the non-linear term cancels the contribution of the real BFKL kernel except in the region $|z-y|<Q_s^{-1}$ in the first term of eq.(\ref{2AEQ2}) and  $|z-x|<Q_s^{-1}$ in the second term of eq.(\ref{2AEQ2}). These leftovers are almost cancelled by the appropriate part of the virtual integral. The remainder depends on the  small size dipoles, so that it plays the role of a source in the equation:
\begin{eqnarray}\label{518r}
-\frac{\partial \bar n(x,y;\eta)}{\partial\,\eta}\,\,&=&-\frac{\bas}{
\pi}\,\int_{|z-x|>Q_s^{-1},|z-y|>Q_s^{-1}}\,K\left(x,y|z
\right)\,\bar n(x,y;\eta)\, \\
&+&\frac{\bas}{2
\pi}\,\left[\int_{|z-x|<Q_s^{-1}}\,K\left(x,y|z
\right)\, \bar n(x,z;\eta) \,+\int_{|z-y|<Q_s^{-1}}\,K\left(x,y|z
\right)\,\bar n(z,y;\eta)\right]\nonumber
\end{eqnarray}
The only difference between this equation and eq.(\ref{tilden}) is the last line. Without this term the density $\bar n$ decreases towards small $\eta$, as discussed above.
The source term itself is positive and thus potentially could change this behavior.  The equation can be re-written in the form:
\begin{eqnarray}\label{5182}
-\frac{\partial \bar n(x,y;\eta)}{\partial\,\eta}\,\,&=&-\bas \,\ln \Lb(x - y)^2 Q^2_s\Lb \eta\Rb \Rb\,\,\bar n(x,y;\eta)\,+\,\bas\,\,\int^{Q^{-2}_s\Lb \eta\Rb}_0\frac{ d (x - z)^2}{(x - z)^2}\, \bar n(x -z;\eta) 
\end{eqnarray}
To understand whether  the source term can have an important effect, we have to consider the evolution of small dipoles as well.

 The evolution of the small dipoles $|x-y|<Q_s^{-1}(\eta)$ is given by 
\begin{eqnarray}
& & - \frac{\partial \bar n(x,y;\eta)}{\partial\,\eta}\,\,= \label{3AEQ2}\\
 &=&\,\,\frac{\bas}{2
\pi}\,\int_z\,K\left(x,y|z
\right)\,\Big\{ \bar n(x,z;\eta) \,+\,\bar n(z,y;\eta)\,-\,\bar n(x,y;\eta)\, \Big\} \nonumber \\
&-& \,\frac{\bas}{ 2\pi}\left\{\int_{|z-x|>Q_s^{-1}(\eta)}\,K \left(x,y|z
\right)\,\ \bar n\left(x,z,\eta\right)+ \int_{|z-y|>Q_s^{-1}(\eta)}\,K \left(x,y|z
\right)\,\bar n\left(z,y,\eta\right)\right\}\nonumber
\end{eqnarray}

Here the last line appears as the source term due to coupling of large dipoles. 
The contribution of large dipoles with sizes greater than the inverse saturation momentum is absent in the RHS of eq.(\ref{3AEQ2}), since the last two terms cancel the contribution of those dipoles to the BFKL kernel. The equation for $\bar n(x-y)$ 
for small dipoles therefore is not sourced by large dipoles.
\begin{eqnarray}
& & - \frac{\partial \bar n(x,y;\eta)}{\partial\,\eta}\,\,= \label{3AEQ31}\\
 &=&\,\,\frac{\bas}{2
\pi}\,\int_z\,K\left(x,y|z
\right)\,\Big\{ \bar n(x,z;\eta)\theta(Q_S^{-1}(\eta)-|x-z|) \,+\,\bar n(z,y;\eta)\theta(Q_S^{-1}(\eta)-|y-z|)\,-\,\bar n(x,y;\eta)\, \Big\} \nonumber 
\end{eqnarray}

It is simple to solve this equation for a certain set of initial conditions. Let us consider  a situation when the unevolved projectile is a single dipole of the size $R_1$ while the target is dense $R_1>Q^{-1}_{s}(0)$, so  that the scattering amplitude on the target is close to unity at all rapidities. For this situation the initial condition for eqs.(\ref{518r},\ref{3AEQ2}) is 
\begin{equation}
\bar n(x-y;Y)=\delta\Big(\ln(x-y)^2/R_1^2\Big)
\end{equation}
 Since the initial dipole size $R_1$ never gets inside the saturation radius throughout the evolution, the small dipole $\bar n$ satisfies at all rapidities a homogeneous equation with the vanishing initial condition. Thus $\bar n(x-y;\eta)=0$ as long as $(x-y)^2Q_s^2<1$.
This also means that there is no source term coming from small dipoles in the equation for large dipoles, and the solution for $\bar n(x-y)$ of arbitrary size is
\begin{equation}
\bar n(x-y)=e^{-\bar\alpha_s \int_{\eta}^Yd\eta \ \ln [R_1^2Q_s^2(\eta)]}\delta\Big(\ln(x-y)^2/R_1^2\Big)
\end{equation}
Although the source term can be important for other initial conditions, it is obvious from the previous discussion that at least for some physically reasonable initial
conditions the introduction of finite $Q_s$ does not change the results of the previous subsection. Namely, it is indeed true that the BK fixed point $(1,0)$ is the attractive fixed point of the Braun Hamiltonian \cite{motyka}. 

In principle, one could perform a similar  more refined analysis of the stability of the point $(1,1)$ allowing for finite $Q_s$ of the projectile and the target. We will not do it here, as the paradoxical nature of the evolution generated by the Braun Hamiltonian is already clear.

\subsection{What does it mean?}
We have established that in the classical approximation to Braun evolution, the Pomeron $\bar \Phi(\eta)$ decreases towards small values of $\eta$, if  $\Phi(\eta)$ is close to unity. This behavior is counterintuitive. Naively $\bar\Phi(\eta)$ has the meaning of the scattering amplitude of a dipole on the projectile wave function at rapidity $Y-\eta$, and so this  seems to suggest that the projectile becomes more transparent to dipoles that have higher energy, if the target is black.

We would now like to be a little less naive and understand more formally what is the meaning of this behavior. 
Solving the classical equations of motion for $\bar P$ is the classical approximation to calculating the rapidity dependent average  $\langle\bar P(\eta)\rangle$. 
 Consider again the correspondence between the CGC expression eq.(\ref{scatmde}) and its path integral representation eq.(\ref{action}). It is clear from this correspondence that the rapidity dependent average of $\bar P$ in the CGC formulation is given by the expression
\begin{eqnarray}
\label{barpcgc}
\langle 1-\bar P(\eta)\rangle&=&\int d\bar P \delta(\bar P)W_P[P]e^{-H_{RFT} [P,\bar P](Y-\eta)}\left( 1-\bar P\right) e^{-H_{RFT} [P,\bar P]\eta}W_T[\bar P]\\
&=&\int d\bar P \delta(\bar P)W_P[P]e^{-H_{RFT} [P,\bar P](Y-\eta)}\left( 1-\bar P\right) W^{\eta}_T[\bar P]\nonumber
\end{eqnarray}
where $W^{\eta}_T[\bar P]$ is the target wave function evolved through the rapidity interval $\eta$. This is the scattering matrix of the projectile on an object which is obtained by evolving the target by rapidity $\eta$, adding to it one extra dipole, and then evolving the resulting system by rapidity $Y-\eta$. What does one expect the $\eta$ dependence of such a scattering matrix to be? Clearly, adding an extra dipole towards the end of the evolution of the target should be less efficient in making the target black than adding it earlier in the evolution. If one adds a dipole early on, it should contribute to subsequent evolution and lead eventually to relatively more dipoles in the wave function, since the QCD evolution always increases the number of physical dipoles. Thus we expect that $\langle1-\bar P(\eta)\rangle$ should increase with $\eta$, and therefore $\bar P(\eta)$ should decrease with $\eta$, or equivalently $\bar P(\eta)$ should increase to smaller values of $\eta$.

As we saw earlier, the behavior of $\bar P$ in the Braun evolution is opposite. There are two possible interpretations of such behavior. One is that the target wave function becomes less dense with evolution. In this case the additional dipole is ``bleached'' by further evolution and contributes little to the scattering amplitude. Physically such behavior is of course completely unacceptable.  The other possibility is that the target does become denser, but the evolution is  so violent that dipoles disappear from its wave function by physically merging with each other. This possibility may be more palatable a priori, however it also is not a part of QCD dynamics. In the leading logarithmic approximation the QCD evolution produces new gluons, and therefore dipoles and never annihilates partons that already exist in the wave function. QCD saturation is the statement that the rate of this growth decreases with the density of the target, but it never completely vanishes and certainly does not become negative.
Thus the behavior of the solutions of the Braun equations indeed violates QCD expectations.

 One could wonder if perhaps this means that the classical approximation to the path integral, which yields this behavior is violated in the dense-dilute regime. It is in principle possible, since the applicability of classical approximation is determined by the magnitude of the sources $J_T$ and $J_P$, and in the dense-dilute limit one of the sources $J_P$ is small. In this case one should be able to see that the loop corrections to the classical approximation are large. However it has been shown in \cite{brauntarasov} that the presence of large $\Phi$ leads to appearance of a kind of "mass" in the Pomeron propagator, and in fact suppresses the loop corrections. The culprit therefore seems to be  the Braun Hamiltonian itself, rather that the classical approximation to the evolution.
 
 We note that this behavior is not unique to the Braun Hamiltonian. In particular we can repeat the same analysis in the framework of the Balitsky-Kovchegov equation, or the dipole model approximation to the JIMWLK Hamiltonian. The result is exactly the same. The evolution of a dense target seems to "bleach" the scattering amplitude of an extra dipole that is added to it. The BK (and JIMWLK) evolution leads to the fixed point $(1,0)$ at large rapidities.

This strange behavior  leads one  to suspect that not all is right with unitarity in the Braun and BK evolution. In particular the evolution of the dense state in any of these frameworks 
may be violating the QCD unitarity. In the next section we will consider this question in the context of a toy model with no transverse dimensions. We will return to the realistic case later.

 \section{Playing with toys I: trouble in the toy world.}
In this section we consider a set of toy models in the framework of the zero transverse dimensional reduction of the RFT\cite{ACJ,AAJ,JEN,ABMC,CLR,CIAF,KLremark2, RS,SHXI,KOLEV,BIT,nestor,LEPRI}. Such models have long been used as a simplified setup for qualitatively understanding of the high energy behavior of QCD. Like in the previous sections we first formulate these models in the framework of the zero dimensional analog of the CGC formalism, and explore their properties.
The scattering matrix of the projectile consisting of $m$ dipoles on the target consisting of $\bar n$ dipoles is given in analogy with the real QCD case by
\beq\label{toyamp}
\langle m|\bar n\rangle=\int d\bar P\delta(\bar P)(1-P)^m(1-\bar P)^{\bar n}
\eeq
This amplitude is evolved in energy according to
\beq
\langle m|\bar n\rangle_Y=\int d\bar P\delta(\bar P)(1-P)^me^{HY}(1-\bar P)^{\bar n}
\eeq
In the following we will consider several toy Hamiltonians. 
\subsection{The BK evolution.}
The zero dimensional analog of the BK evolution\cite{BK} is given by the Hamiltonian\footnote{ Compared to previous sections, 
here we rescale the rapidity variable by the factor $\sqrt{\gamma}\sim \alpha_s$. }
\beq\label{hbk}
H_{BK}=-\frac{1}{\gamma}\left[\bar PP-\bar PP^2\right]
\eeq
As before we take $P$ and $\bar P$ to have the dilute limit algebra, such that
\beq P=-\gamma\frac{d}{d\bar P}; \ \ \ \ \ \gamma \sim \alpha_s^2 >0
\eeq
The constant $\gamma$ is the zero dimensional proxy for the dipole-dipole scattering probability.

The scattering matrix can be calculated explicitly
(we assume $m+1<\bar n$)
\beq
\langle m|\bar n\rangle=\sum_{l=0}^m\frac{m!\bar n!}{(m-l)!(\bar n-l)!l!}(-\gamma)^l
\eeq
In particular
\beq\label{formu}
\langle 1|\bar n\rangle=1-\bar n\gamma
\eeq
It is clear from Eq.(\ref{formu}) that our current formulation is restricted to number of dipoles $\bar n<1/\gamma$, since otherwise the single dipole S-matrix becomes negative. The reason for this unphysical behavior is the dilute limit commutation relation we adopted for $P$ and $\bar P$. As we will see later, with correct commutation relation the problem does not arise. Nevertheless as long as $\bar n<1/\gamma$ we can continue the present analysis.

%
 
 Our aim now is to compare $\langle \bar P(\eta)\rangle$ for two different values of rapidity. For simplicity we will take the rapidity interval to be infinitesimally small, and will simply insert $\bar P$ into the matrix element either before or after this short evolution interval. Also for simplicity we take the projectile to contain a single dipole, although the calculation can be easily generalized. Thus we are interested in
 \begin{equation}
 \langle 1-\bar P\rangle_P\equiv \langle 1|(1-\bar P)e^{H\Delta}|\bar n\rangle; \ \ \ \ \ \ \  \langle 1-\bar P\rangle_T\equiv \langle 1|e^{H\Delta}(1-\bar P)|\bar n\rangle
 \end{equation}
 As discussed in the previous section, the two quantities have simple physical meaning. The first one corresponds to the evolution of the target wave function with subsequent insertion of an additional dipole, prior to scattering on the projectile. In the second one we insert an extra dipole into the target wave function, then evolve the combined target+dipole, and then scatter it on the projectile. The physical expectation is that inserting the extra dipole closer to the target will produce a blacker target. Thus we expect
 \begin{equation}\label{comp}
  \langle 1-\bar P\rangle_P> \langle 1-\bar P\rangle_T\ \ \ ?
  \end{equation}
  As we will see, this expectation is fulfilled when $\bar n\sim 1$. However the evolution produces the opposite result when $\bar n\gamma\sim 1$, that is for dense target.
  
 For comparison we will also calculate an analogous quantity for  the pomeron $P$
  \begin{equation}
   \langle 1-P\rangle_P\equiv \langle 1|(1- P)e^{H\Delta}|\bar n\rangle; \ \ \ \ \ \ \  \langle 1-P\rangle_T\equiv \langle 1|e^{H\Delta}(1-P)|\bar n\rangle
 \end{equation}
 The interpretation of these quantities is the dual of those discussed above. Therefore we expect 
  \begin{equation}\label{comp1}
  \langle 1- P\rangle_P< \langle 1-P\rangle_T\ \ \ ?
  \end{equation}  
  For the pomeron $P$ we obtain
\begin{eqnarray}\label{pomerons}
\langle 1-P\rangle_P&=&=\left[1-2\bar n\gamma+\bar n(\bar n-1)\gamma^2\right]+2\Delta\left[-\bar n\gamma+2\bar n(\bar n-1)\gamma^2-\bar n(\bar n-1)(\bar n-2)\gamma^3\right]\\
\langle 1-P\rangle_T&=&\left[1-2\bar n\gamma+\bar n(\bar n-1)\gamma^2\right]+\Delta\left[-\bar n\gamma+2\bar n(\bar n-1)\gamma^2-\bar n(\bar n-1)(\bar n-2)\gamma^3\right]\\
\langle 1-P\rangle_P&-&\langle 1-P\rangle_T=-\Delta\bar n\gamma\Big\{\left[1-(\bar n-1)\gamma\right]^2-(\bar n-1)\gamma^2\Big\}<0
\end{eqnarray}
where the last equality holds for $\bar n<1/\gamma$.
Thus the behavior of $P$ conforms with our expectations.
The same calculation for the conjugate Pomeron $\bar P$ yields
\begin{eqnarray}\label{barpomerons}
  \langle 1-\bar P\rangle_P&=&1-(\bar n+1)\gamma -\Delta\bar n\gamma\Big[1-(\bar n-1)\gamma\Big]\\
\langle 1-\bar P\rangle_T&=&1-(\bar n+1)\gamma-\Delta(\bar n+1)\gamma\Big[1-\bar n\gamma\Big]\nonumber\\
  \langle 1-\bar P\rangle_P&-& \langle 1-\bar P\rangle_T=\Delta\gamma \Big[1-2\bar n\gamma\Big]
  \end{eqnarray}
  Thus for small $\bar n$ eq.(\ref{comp}) is satisfied, however for $\bar n>\frac{1}{2\gamma}$ the situation is reversed.   
  The toy model thus behaves in the similar way to the model with two transverse dimensions.

To understand the origin of this behavior consider the infinitesimal evolution of the projectile and target wave functions with the Hamiltonian $H$:
\begin{equation}\label{projevolv}
\langle m|e^{\Delta H}\approx  (1-\Delta m )\langle m|+\Delta m \langle m+1|
 \end{equation}
 \begin{equation}\label{tarevolv}
  e^{\Delta H}|\bar n\rangle=(1+\Delta\bar n)|\bar n\rangle -\Delta\bar n[1+\gamma(\bar n-1)]|\bar n-1\rangle+\Delta\gamma\bar n(\bar n-1)|\bar n-2\rangle
 \end{equation}
 Note the sea of difference between the two expressions. Recall that the coefficient in front of an $n$-dipole state has the meaning of probability to find this number of dipoles in the the wave function. The projectile evolution is unitary: all the probabilities in the evolved state remain positive and smaller than unity, and the sum of the probabilities adds up to unity.
 
  On the other hand the target evolution is non-unitary. The probability to find the initial state $|\bar n\rangle$ after a short interval of evolution exceeds unity, while the probability to find a state with one less particle is negative. The coefficients still sum to unity like for the projectile, but clearly  the target evolution violates unitarity.

  We stress that the probabilities in question are 
 probabilitites to find physical dipoles in the wave function of the evolved hadronic state. Thus the negativity of probabilities violates the $s$ - channnel unitarity. This violation is not directly seen in the calculation of the diagonal matrix element of the $S$-matrix, since by construction this matrix element is  a real number smaller than one. However it is clear that if we were to consider  more exclusive observabes, the negativity of probabilities would  show up as unphysical values for some observable. Although a detailed study of this question is outside the scope of this paper, it is easy to give an example of such  an observable. Consider  a  target containing $n$ dipoles, which  scatters on a dense projectile. If the projectile is very dense, all the dipoles in the target wave function will be scattered into the final state. Thus the probability for producing $n$ dipoles in the final state will be equal to the probability of finding $n$ dipoles in the wave function. At a slightly higher energy  than the initial one, the probability of producing $n-1$ dipoles in the final state will be negative.
 Strictly speaking those are of course   "toy dipoles" in the "toy hadron", but the essence of the argument is the same in real QCD.

 It is interesting to examine more carefully the evolved target side wave functions that enter in the calculations in eq.(\ref{barpomerons}) at the rapidity they scatter on the projectile dipole.
 \begin{eqnarray}
&& e^{\Delta H}(1-\bar P)|\bar n\rangle = \Big[1+\Delta(\bar n+1)\Big]|\bar n+1\rangle -\Delta(\bar n+1)[1+\gamma\bar n]|\bar n\rangle+\Delta\gamma(\bar n+1)\bar n|\bar n-1\rangle \\
 &&(1-\bar P)e^{\Delta H}|\bar n\rangle  =\Big[1+\Delta\bar n\Big]|\bar n+1\rangle -\Delta\bar n[1+\gamma(\bar n-1)]|\bar n\rangle+\Delta\gamma\bar n(\bar n-1)|\bar n-1\rangle \end{eqnarray}
 Calculating the average number of dipoles in the two wave function we find
 \begin{eqnarray}\label{numbers}
 &&\langle N\rangle_T=\bar n+1+\Delta(\bar n+1)-\Delta\gamma(\bar n+1)\bar n\\
 &&\langle N\rangle_P=\bar n+1+\Delta\bar n-\Delta\gamma(\bar n-1)\bar n\\
 &&\langle N\rangle_T-   \langle N\rangle_P=\Delta\Big[1-2\gamma\bar n\Big]
 \end{eqnarray}
 This indeed displays the poignant feature discussed above, namely at small $\bar n$ the average number of dipoles in the wave function is larger if an extra dipole is added before evolution, while at large $\bar n>1/2\gamma$ the situation is reversed.
The reason for  negative difference in eq.(\ref{numbers}) is obvious. It appears because the negative probability of the $|\bar n-1\rangle$ contribution in eq.(\ref{tarevolv}) grows with $\bar n$ faster than the positive probability of the $|\bar n\rangle$ contribution. 

Thus we see that the nonunitarity of the BK evolution {\it of the target wave function} is indeed the reason for the counter intuitive behavior of $\bar P$ with rapidity.

To summarize, we have shown that although the (zero dimensional) BK-JIMWLK evolution of the projectile wave function preserves QCD unitarity, the same evolution when viewed as evolution of the target wave function is non-unitary. 
 
 \subsection{The Braun Hamiltonian.}
 
 Next consider the analog of the Braun Hamiltonian
 \begin{equation}\label{rbaun}
 H_B=-\frac{1}{\gamma}\left[\bar PP-\bar PP^2-\bar P^2P\right]
 \end{equation}
We pose the same question: does this Hamiltonian generate a unitary evolution? To answer this we consider, as before
\begin{equation}\label{tarevolb}
e^{\Delta H_B}|\bar n\rangle\approx (1+\Delta H)|\bar n\rangle=(1-\Delta\bar n)|\bar n\rangle +\Delta\bar n|\bar n+1\rangle-\Delta\gamma\bar n(\bar n-1)]|\bar n-1\rangle+\Delta\gamma\bar n(\bar n-1)|\bar n-2\rangle
\end{equation} 
This is somewhat more satisfactory than eq.(\ref{tarevolv}), since the violation of unitarity is $O(\gamma)$ and is small for small $\bar n$ . However the coefficient of the term $|\bar n\rangle$ is still negative, and becomes large parametrically long before the saturation limit is reached. 
Alarmingly, since the Braun evolution is symmetric between the target and the projectile, the projectile evolution now is also non-unitary and involves negative probabilities.

We note, that the Braun  eq.(\ref{rbaun}) has been considered in the past from the point of view of the reaction-diffusion process (RDP)\cite{shoshi}. Ref. \cite{shoshi} indeed made it explicit that this evolution corresponds to a non-unitary RDP that involves negative emission probabilities. The RDP emission probabilities are however distinct from the QCD probabilities and in fact not related to them in a simple obvious way. Thus the violation of unitary we discuss here is distinct from, and not obviously related to the nonunitarity of the appropriate RDP.

There may be more than one problem in the previous models. In particular we have seen that the commutator we have postulated between $P$ and $\bar P$ can only be used for a target with small enough number of dipoles, otherwise even without any evolution the $S$-matrix is non-unitary. In particular $\langle 1|\bar n\rangle<0$ for large enough $\bar n>1/\gamma$. One could perhaps wonder if this deficiency is to blame for the nonunitarity of the evolution as well. In the rest of this this section we will rectify this deficiency and show how to define the correct commutation relation. We will also show that even with the redefined commutation relation, the BK and Braun Hamiltonians lead to non-unitary evolution.

\subsection{Are the commutators to blame?}
Our postulated commutation relation does not allow for multiple scattering corrections when a single dipole of the projectile scatters on several dipoles of the target.
 Clearly the correct formula for scattering of one dipole on $\bar n$ dipoles should be
\begin{equation}
\langle 1|\bar n\rangle=\sum_{k=0}^{\bar n}\frac{\bar n!}{(\bar n-k)!k!}(-\gamma)^k
\end{equation}
since this expression correctly accounts for multiple scattering corrections. 
The algebra of $P$ and $\bar P$ should be such that this result follows from the definition of the amplitude eq.(\ref{toyamp}). 

A simple way to achieve this is to modify the relation between $P$ and $\bar P$ as follows 
\begin{equation}
P=1-e^{\gamma\frac{d}{d\bar P}}\  \  \  \  \ ?
\end{equation}
This is better, but still not good enough. In particular it does not allow two dipoles of the projectile to scatter on the same dipole of the target, since the first factor $1-P$ by differentiation simply kills the particular target dipole, and subsequent scatterings on it are not possible. The propagation of the projectile dipole should be ``non demolition'', in the sense that after moving the factor $1-P$ through $\bar P$, the factor $\bar P$ should not disappear from the wave function $W_T$. Therefore a more reasonable representation is
\begin{equation}\label{pdem}
1-P=\sum_{k=0}^\infty \frac{1}{k!}\gamma^k(1-\bar P)^k\frac{d^k}{d\bar P^k}\  \  \  \  \  \  ?
\end{equation}
However eq.(\ref{pdem}) is not quite adequate either. According to it the propagation of the projectile dipole does not destroy any target dipoles, but the projectile dipole itself disappears after propagation, and this is not right. 

None  of the above problems arise if the $P$ and $\bar P$ algebra is taken to be the following 
\beq\label{ccom}
(1-P)(1-\bar P)=[1-\gamma](1-\bar P)(1-P)
\eeq
This ensures that moving one projectile dipole through $\bar n$ target dipoles give the correct factor $(1-\gamma)^{\bar n}$ that includes all multiple scattering corrections, while all the dipoles remain intact, and can subsequently scatter on additional projectile or target dipoles.
For small $\gamma$  and in the regime where  $P$ and $\bar P$ are small themselves, we obtain
\beq
[P,\bar P]=-\gamma +...
\eeq
consistently with our original expression (\ref{commut},\ref{barphi}). 

Note that the algebra eq.(\ref{ccom}) is equivalent to the following representation
\beq \label{defin} 1-\bar P=e^{ -\ln(1-\gamma)\frac {d}{d\Phi}}, ; \ \ \ \ \ \ 1-P=e^{-\Phi}
\eeq
In the calculation of an amplitude of the type of eq.(\ref{toyamp}), once all the factors of $1-\bar P$ are commuted through to the left, in any matrix element $\bar P$ hits the $\delta$-function and thus vanishes.
The remaining factors of $(1-P)$ also turn to unity, since a  factor of $\Phi$ is equivalent to a derivative acting on the $\delta$-function, and when integrated over $\bar P$ vanishes.

 With the new algebra we have
\begin{equation}\label{unit}
\langle m|\bar n\rangle=(1-\gamma)^{m\bar n}
\end{equation}
which is a simple and intuitive result: the s-matrix of dipole-dipole scattering to the power of the number of dipole pairs that scatter.

We stress that the modification of the Pomeron algebra is not a matter of choice, but is necessary to obtain the amplitude eq.(\ref{unit}), which is unitary for arbitrary numbers of colliding dipoles. However the question of the unitarity of the evolution is a completely separate one. We will now reexamine the BK and Braun evolutions with the modified Pomeron algebra.

\subsection{BK evolution revisited: the Hamiltonian with modified commutators.}

We start with the BK Hamiltonian defined in eq.(\ref{hbk}). As before we ask if evolution by the infinitesimal rapidity interval preserves the probabilistic interpretation of the initial wave function.
\begin{equation}\label{tarevbk}
e^{\Delta H_{BK}}|\bar n\rangle=\Big[1-\frac{\Delta}{\gamma}\left[1-(1-\gamma)^{\bar n}\right](1-\gamma)^{\bar n}\Big]|\bar n\rangle+\frac{\Delta}{\gamma}\left[1-(1-\gamma)^{\bar n}\right](1-\gamma)^{\bar n}|\bar n+1\rangle
\end{equation}
\begin{equation}
\langle m|e^{\Delta H_{BK}}
=\langle m|-\frac{\Delta}{\gamma}\left[1-(1-\gamma)^m\right]\langle m+1|+
\frac{\Delta}{\gamma}\left[1-(1-\gamma)^m\right]\langle m+2|
\end{equation}
This result is surprising:
the evolution of the {\it target} is now unitary, but of the projectile is not. The evolution on the target wave function looks reasonable. When $\bar n$ is small, it is identical with the BFKL evolution. For large $\bar n$ it  exhibits very strong saturation due to suppression with the factor $(1-\gamma)^{\bar n}$, so that at large $\bar n$ the evolution is super slow. This is a little disturbing, but does not seem fatal. However the projectile now evolves nonunitarity, and thus we expect the same type of trouble as found in the previous subsection.

Let us see how this reflects on the behavior of $P$ and $\bar P$.
Calculating  $\langle P\rangle $ we find
\begin{eqnarray}
\langle 1-P\rangle_P&=&(1-\gamma)^{2\bar n}-\frac{\Delta}{\gamma}\left[1-(1-\gamma)^2\right](1-\gamma)^{3\bar n}+\frac{\Delta}{\gamma}\left[1-(1-\gamma)^2\right](1-\gamma)^{4\bar n}\\
\langle 1-P\rangle_T&=&(1-\gamma)^{2\bar n}-\frac{\Delta}{\gamma}\left[1-(1-\gamma)\right](1-\gamma)^{3\bar n}+\frac{\Delta}{\gamma}\left[1-(1-\gamma)\right](1-\gamma)^{4\bar n}\nonumber
\end{eqnarray}
So that 
\beq
\langle 1-P\rangle_P-\langle 1-P\rangle_T=-\Delta (1-\gamma)^{3\bar n+1}\Big[1-(1-\gamma)^{\bar n}\Big]<0
\eeq
This difference is always negative, consistently with our logic, even though the evolution as we saw is nonunitary.

Now for $\langle \bar P\rangle$:
\begin{eqnarray}
\langle 1-\bar P\rangle_P&=&
\Big[1-\frac{\Delta}{\gamma}\left[1-(1-\gamma)^{\bar n}\right](1-\gamma)^{\bar n}\Big](1-\gamma)^{\bar n+1}+\frac{\Delta}{\gamma}\left[1-(1-\gamma)^{\bar n}\right](1-\gamma)^{\bar n}(1-\gamma)^{\bar n+2}\nonumber\\
\langle 1-\bar P\rangle_T&=&
\Big[1-\frac{\Delta}{\gamma}\left[1-(1-\gamma)^{\bar n+1}\right](1-\gamma)^{\bar n+1}\Big](1-\gamma)^{\bar n+1}+\frac{\Delta}{\gamma}\left[1-(1-\gamma)^{\bar n+1}\right](1-\gamma)^{\bar n+1}(1-\gamma)^{\bar n+2}\nonumber
\end{eqnarray}
So that 
\begin{equation}
\langle 1-\bar P\rangle_P-\langle 1-\bar P\rangle_T
=\Delta\gamma(1-\gamma)^{2\bar n+1}\Big[(2-\gamma)(1-\gamma)^{\bar n}-1\Big]\nonumber
\end{equation}
This has the same behavior as before. For small $\bar n$ this difference is positive, while for $\bar n>-\frac{\ln (2-\gamma)}{\ln (1-\gamma)}$ it changes sign, and thus it again manifests nonunitarity of the evolution.    

\subsection{The Braun Hamiltonian with modified commutators.}
Next let us examine the evolution generated by the Braun Hamiltonian.
 The analog of the original Braun Hamiltonian eq.(\ref{braun}) is 
 \begin{equation}\label{1rbraun}
\tilde H_B=-\frac{1}{\gamma}\left[\bar \Phi\Phi-\bar\Phi\Phi^2-\bar\Phi^2\Phi\right]
 \end{equation}
with $\Phi=\ln (1-P);\ \ \bar\Phi=\ln (1-\bar P)$. The variables $\Phi$ and $\bar \Phi$ are canonically conjugate. The modified commutation relation between $P$ and $\bar P$ however enters in the calculation of matrix elements, as the projectile and target wave functios carry factors of $(1-P)^n; \ \ (1-\bar P)^{\bar n}$, see eqs.(\ref{wp},\ref{wt}). The action of this  Hamiltonian is obviously nonunitary, since a basic necessary condition is that the coefficients in the expansion of $H$ in powers of $(1-P)$  be finite. The logarithmic factors in eq.(\ref{1rbraun}) obviously yield infinite expasion coefficients. However using instead the form eq.(\ref{rbaun}), which is self dual and reduces to eq.(\ref{1rbraun}) in the dilute limit solves this problem. This form of the Braun Hamiltonian has a chance to be unitary, and we will concentrate on this question.  

We rewrite the Braun Hamiltonian eq.(\ref{rbaun}) in a more convenient form: 
\beq
H_B=-\frac{1}{\gamma}\Big[(1- \bar P)P-(1-\bar P)^2P+(1-\bar P)P^2-P^2\Big]
\eeq
The action on the projectile and the target is obviously symmetric, as the Hamiltonian is self dual under the transformation $P\rightarrow \bar P$.
\beq
e^{\Delta H_B}|\bar n\rangle=\left[1+\frac{\Delta}{\gamma}\left[1-(1-\gamma)^{\bar n}\right]^2\right]|\bar n\rangle-\frac{\Delta}{\gamma}\left[1-(1-\gamma)^{\bar n}\right]\left[2-(1-\gamma)^{\bar n}\right]|\bar n+1\rangle+\frac{\Delta}{\gamma}\left[1-(1-\gamma)^{\bar n}\right]|\bar n+2\rangle
\eeq
\beq
\langle m|e^{\Delta H_B}=\left[1+\frac{\Delta}{\gamma}\left[1-(1-\gamma)^{m}\right]^2\right]\langle m|-\frac{\Delta}{\gamma}\left[1-(1-\gamma)^{m}\right]\left[2-(1-\gamma)^{m}\right]\langle m+1|+\frac{\Delta}{\gamma}\left[1-(1-\gamma)^{m}\right]\langle m+2|
\eeq
This is a nasty surprise. Now the evolution of both, projectile and target is non-unitary. In fact the lack of unitarity is there for arbitrary 
number of dipoles $m$ and $\bar n$.

We may hope that modifying the Braun Hamiltonian with an extra $\bar P^2P^2$ term could improve the situation.
However, it does not bring about complete redemption. Consider
\beq\bar H_B=H_B-\frac{1}{\gamma}\bar P^2P^2=-\frac{1}{\gamma}\left[ \bar PP-\bar PP^2-\bar P^2P+\bar P^2P^2\right]=-\frac{1}{\gamma}\left[(1-\bar P)-(1-\bar P)^2\right][P-P^2]
\eeq
Now we obtain:
\beq
e^{\Delta \bar H_B}|\bar n\rangle=|\bar n\rangle-\frac{\Delta}{\gamma}\left[1-(1-\gamma)^{\bar n}\right](1-\gamma)^{\bar n}\Big\{|\bar n+1\rangle-|\bar n+2\rangle\Big\}
\eeq 
\beq
\langle m|e^{\Delta \bar H_B}=\langle m|-\frac{\Delta}{\gamma}\left[1-(1-\gamma)^{m}\right](1-\gamma)^{m}\Big\{\langle m+1|-\langle m+2|\Big\}
\eeq 
  Unfortunately this is as non-unitary as the BK evolution with the slight modification of the approach to saturation. 
It is not difficult to show that the nonunitarity cannot be cured by adding the four Pomeron vertex with any coefficient.

\section{Playing with toys II: making the toy world a better place.}
It may seem that our modification of the commutation relations was in vain as it did not solve the problem of non-unitary evolution. However, as it happens often, good deeds get rewarded. 
In this section using the correct commutation relations we will be able to find a Hamiltonian which has the correct dense-dilute limit, is self dual and produces unitary evolution of both, the projectile and the target.

\subsection{Unitarity regained.}
The discussion of the previous section does not necessarily mean that we are doomed to live with non-unitary evolution.
There is one thing that we have so far implicitly accepted, namely the form of the BK/Braun hamiltonian in terms of the Pomeron operators. This is so even though we have not derived it directly. What is derivable from QCD is the Hamiltonian in terms of $P$ and $P^\dagger$, rather than $P$ and $\bar P$.
Since $P^\dagger$ and $\bar P$ are only proportional to each other in the dilute limit, our use of the BK and Braun Hamiltonians away from this limit is not justifiable.
We do know however, that  the correct unitary Hamiltonian (if it exists) has to reduce to the $H_{BK}$ in the limit of small $\bar P$.
We will now attempt to modify the BK Hamiltonian in a way that makes it unitary, but still reduces to the original $H_{BK}$ when expanded to linear order in $\bar P$.

First, we express the $P^\dagger$ in terms of $\bar P$.
To do this recall that $P^\dagger$ should annihilate a dipole when acting on the wave function. Using eq.(\ref{defin}) we can write
\beq \label{daggerbar}
P^\dagger= \frac{d}{d\Phi} e^{\Phi}=\frac{1}{\gamma}\ln(1-\bar P)\frac{1}{1-P}; \ \ \ \ \ \  \bar P^\dagger=-\frac{1}{\gamma} e^{-\gamma\frac{d}{d\Phi}}\Phi=\frac{1}{\gamma}\frac{1}{1-\bar P}\ln(1-P)\eeq
Thus the BK Hamiltonian  expressed in terms of $P$ and $\bar P$ is
\begin{equation} \label{HX}
H_{BK}=P^\dagger \left[P-P^2\right]=\frac{1}{\gamma}\ln (1-\bar P)P
\end{equation}
We can also conveniently write its dual (the mean field approximation to the KLWMIJ Hamiltonian) as
\begin{equation}\label{KB}
H_{KB}=\left[\bar P-\bar P^2\right]\bar P^\dagger=\frac{1}{\gamma}\bar P\ln (1-P)
\end{equation}
In the above equations for simplicity we have used $\ln(1-\gamma)\approx -\gamma$, since $\gamma\sim \alpha_s^2\ll 1$.

To define the Braun Hamiltonian we have to add these two and subtract the BFKL limit. The simplest analog of the BFKL Hamiltonian is  
the leading order expansion of either one of eq.(\ref{HX}) or eq.(\ref{KB}) in $\Phi$ and $d/d\Phi$
\beq \label{HX0}
H_{BFKL}^1=-\frac{d}{d\Phi}\Phi=-\frac{1}{\gamma}\ln (1-\bar P)\ln (1-P); 
\eeq

We thus can write an analog of Braun Hamiltonian as
\beq\label{HB1} H_B^1=\frac{1}{\gamma}\left[ \ln (1-\bar P)P+\bar P\ln (1-P)+\ln (1-\bar P)\ln (1-P)\right]\eeq
This Hamiltonian is clearly non-unitary. One does not need to perform any calculation to understand this. Our unitarity test of the projectile  evolution amounts to the following simple three step procedure: 

1. Act with the Hamiltonian on a monomial $(1-P)^n$ ; 

2. Expand the result in powers of $(1-P)$; 

3. Check that the coefficients of all terms $(1-P)^m;\ \ m\ne n$ are positive, and the coefficient of $(1-P)^n$ is negative.

For the target the same procedure is applied to  $(1-\bar P)^n$.

This set of conditions can be formulated as the following requirements on the Hamiltonian. Write the Hamiltonian as a function of $d=1-P$ and $\bar d=1-\bar P$; $H(d,\bar d)$. When the hamiltonian is commuted through $\bar d^n$ to the right, each operator $d$ turns into one-on-$n$ scattering amplitude,  a positive number smaller than one, which we can also denote as $d$. Thus our requirements can be written as 
\begin{eqnarray}&& H(d, \bar d=0)<0; \ \ \ \ \ \ \frac{\partial^k}{\partial \bar d^k}H(d,\bar d)|_{\bar d=0}>0\, \ \ \ \ \ {\rm for \,\, any \,} d:\, \  0<d<1; \ k\ge 1 \\ 
&&      H(d=0; \bar d<0 )         ; \ \ \ \ \ \ \frac{\partial^k}{\partial d^k}H( d, \bar d)|_{ d=0}>0\, \ \ \ \ \ {\rm for \,\, any \,} \bar d:\, 
\  0<\bar d<1; \ k\ge 1\nonumber
\end{eqnarray}

It is obvious that neither $H_B^1$, nor $H_{BK}$ nor $H_{KB}$ is unitary, since they all contain logarithmic factors. Thus step 2 in our procedure fails, as it gives infinite coefficients. Equivalently, the derivatives of the Hamiltonian at $\bar d=0$ diverge.

However the proposal eq.(\ref{HB1}) for the Braun Hamiltonian is not unique. It was written on the basis of two requirements: it should be self dual under $P\leftrightarrow \bar P$; and for small $\bar P$ ($P$) it should reduce to $H_{BK}$ ($H_{KB}$). It is in fact possible to write down a Hamiltonian that satisfies these requirements, as well as the unitarity constraint:


\begin{equation}\label{hb}
H_{UTM}=-\frac{1}{\gamma}\bar PP\eeq
where $UTM$ stands for ``Unitarized Toy Model''.
The fact that it is self-dual is evident. 
Expanding $\bar P$ to linear order in $P^\dagger$, using eq.(\ref{daggerbar}) leads to the BK Hamiltonian, eq.(\ref{HX}). 

To check the unitarity we consider:
\beq \label{utmu}e^{\Delta H_{UTM}}|\bar n\rangle=\left[1-\frac{\Delta}{\gamma}[1-(1-\gamma)^{\bar n}]\right]|\bar n\rangle+\frac{\Delta}{\gamma}[1-(1-\gamma)^{\bar n}]|\bar n+1\rangle
\eeq
This evolution is clearly unitary. Due to self duality, it is clear that the evolution of the projectile wave function is unitary as well.
 Interestingly it also exhibits the saturation behavior very similar to the one that is expected from the real QCD evolution, namely at large $\bar n$, the change in the wave function is independent of the number of dipoles $\bar n$. In this respect it contrasts strongly with eqs.(\ref{projevolv}) and (\ref{tarevbk}), which were also unitary. In the standard BK evolution of the projectile eq.(\ref{projevolv}) the wave function never saturates, meaning the rate of growth of number of dipoles is proportional to the number $m$ of dipoles in the state even for very large $m$. This is of course the well known property of the BK evolution, where the projectile state evolves according to the perturbative dipole model and saturation of the scattering amplitude is only due to the multiple scattering effects. Eq. (\ref{tarevbk}) on the other hand is completely different. Its evolution is "oversaturated", in the sense that for large $\bar n$, the wave function does not evolve at all. Clearly this cannot be a reflection of a QCD-like dynamics.

An interesting and very appealing property of the UTM Hamiltonian, is that one can arrive at it either from BK by expanding $P^\dagger$ to leading order in $\bar P$, or from KB by expanding $\bar P^\dagger$ to leading order in $P$, or indeed from BFKL by using both expansions. 

Having found a unitary evolution it is interesting to explore its properties. In the next subsection we provide a solution of the classical equations that follow from $H_{UTM}$.

\subsection{Equations of motion and the scattering amplitude.}

 The general form of equation of motion follows from
 \beq \label{EM1}
 \frac{d P}{d  \eta}\,=\,\Big[ H, P\Big]\,;~~~~{\rm and}~~~~ 
 \frac{d \bar P}{d  \eta}\,=\,\Big[ H, \bar P\Big]
\eeq
With the Hamiltonian $H_B$ we get 
\beq \label{H01}
\frac{ d P}{d  \eta}\,=\,\Lb 1 - \bar P\Rb \,\Lb 1 - P \Rb \,P;~~~~~~~\frac{ d \bar P}{d  \eta}\,=\,-\Lb 1 - P\Rb \,\Lb 1 - \bar P \Rb \,\bar P;
\eeq
Interestingly, although it is not obvious from the form of the Hamiltonian Eq.(\ref{hb}), the evolution has the same fixed points as in two transverse dimensions $(0,0), \ (1,0), \ (0,1), \ (1,1)$.

Since the Hamiltonian is conserved, we have 
\beq\label{con}
\bar P P = {\rm Const} (\eta) \equiv {\rm \alpha}
\eeq
The general solution to \eq{H01} takes the form:
 \beq \label{H03}
 P (\eta)\,=\,\frac{ \alpha +\beta e^{ (1 - \alpha) \eta} }{1 +\beta e^{ (1 - \alpha)  \eta}}; \ \ \ \ \bar P(\eta)=   \frac{ \alpha \Big(1+\beta e^{ (1 - \alpha) \eta}\Big) }{\alpha +\beta e^{ (1 - \alpha)  \eta}};
 \eeq
 where the parameters $\beta$ and $\alpha$ should be found from the boundary conditions:
 \beq \label{H0BC}
 P (\eta= 0)\,=\,p_0;\,\,\,\,\,\,\,\, \bar P (\eta= Y)\,=\,\frac{\alpha}{P (\eta= Y)}\,=\,\bar p_0
 \eeq
 One can see that for $p_0 \,>\,\bar p_0$ and $e^{(1 - \alpha)Y}\,\gg\,1$,  \eq{H0BC} leads to
 \beq \label{H031}
 \beta\,=\,\frac{p_0 \,-\,\alpha}{1\,-\,p_0}\,=\,\frac{p_0 \,-\,\bar p_0}{1 - p_0}; \,\,\,\,\,\,\,\,\,\,\alpha\,=\,\bar p_0;
 \eeq 
 For  a symmetric boundary condition $p_0 = \bar p_0$ \eq{H0BC} give $P(0)=\bar P(Y)$ and the solution takes the form
 
  
\begin{eqnarray}
P(\eta)&=&\frac{\alpha+\sqrt{\alpha}e^{(1-\alpha)(\eta-Y/2)}}{1+\sqrt{\alpha}e^{(1-\alpha)(\eta-Y/2)}}\\
\bar P(\eta)&=&\frac{\alpha\left(1+\sqrt{\alpha}e^{(1-\alpha)(\eta-Y/2)}\right)}{\alpha+\sqrt{\alpha}e^{(1-\alpha)(\eta-Y/2)}}\nonumber\\
P(\eta)&=&\bar P(Y-\eta)\nonumber
\end{eqnarray}
This solution has a distinct BFKL-like regime. 
Let us take $\alpha\ll 1$ and $e^{-Y/2}=a\sqrt{\alpha}$ with $1/\alpha\gg a\gg 1$. 
We then have
\beq P(\eta)\approx a\alpha e^{\eta}\eeq
 The exponential ``BFKL'' growth continues until the Pomeron reaches the value $P(Y)=1/(1+a)$.

In \fig{solh0} we have plotted  numerical  solutions to \eq{H01} that correspond to different initial conditions. The BFKL-like regime is clearly seen on \fig{solh0}-a.
All the solutions clearly show that $P$ grows towards positive rapidities, while $\bar P$ grows towards negative rapidities. This is of course a direct consequence of the conservation of $\bar PP$, and thus the unitarized evolution indeed cures the peculiarity of the evolution of $\bar P$. 

However we learn from these solutions that our initial expectation that for large $Y$ at intermediate rapidities the solution should be dominated by the fixed point $(1,1)$ is not warranted. Although both $P$ and $\bar P$ grow towards midrapidity, there is no value of rapidity at which they are simultaneously close to unity, unless it is forced by the initial conditions. 

In fact, once we account for the conservation of $\bar PP$, we get a very different view of the fixed point structure of the evolution. 
Plugging the relation Eq.(\ref{con}) in \eq{H01} we obtain 
 the following equations:
 \beq \label{H011}
 \frac{ d P}{d  \eta}\,=\,\Lb  P\,-\,\alpha \Rb \,\Lb 1 - P \Rb ; \ \ \ \ \ \  \ \ \frac{ d \bar P}{d  \eta}\,=-\,\Lb  \bar P\,-\,\alpha \Rb \,\Lb 1 - \bar P \Rb 
 \eeq
 
 The fixed points $(0,0),\ (0,1)$ and $(1,0)$ are not present in these equations, which means that neither one of them is reachable at $\alpha\ne 0$. The point $(1,1)$ is also unreachable by the evolution for $\alpha\ne 1$.
 \eq{H011} has  only two interesting fixed points: $\Lb 1,\alpha\Rb$ and $\Lb \alpha, 1\Rb$. Since for any physical initial condition $P(0)>\alpha$, the asymptotics at $\eta\rightarrow\infty$ is always dominated by the fixed point $(P=1,\bar P=\alpha)$, while for $\eta\rightarrow 0$ the point $(P=\alpha, \bar P=1)$ is approached. Whether either one of these points is reached during the evolution to finite $Y$ depends on the initial conditions. As illustrated on \fig{solh0}-a,b for symmetric initial condition the solution approaches very close to the fixed points at both ends, while for asymmetric initial conditions this is not the case, and only the vicinity of one fixed point is reached. In \fig{solh0}-c-f this is the point $(1,\alpha)$ at $\eta\rightarrow Y$. There are of course mirror solutions where instead the point $(\alpha,1)$ is approached at $\eta\rightarrow 0$.

Another noteworthy property of the solution is, that for strongly asymmetric initial conditions, the smaller of the two Pomerons remains small essentially over the whole evolution. This is clearly seen in \fig{solh0}-d. The physical reason for that is the saturation effects in the wave function. As we have seen in eq.(\ref{utmu}), when the target wave function contains many dipoles  ($P$ is close to unity), the rate of increase of the dipole number is constant and independent  on the number of dipoles . Consider the dependence of $\bar P$ on $\eta$. As we have discussed in detail in the previous sections, the classical solution for $1-\bar P(\eta)$ is  the scattering amplitude on the projectile of the target with an extra dipole inserted  at rapidity $\eta$. Inserting an extra dipole at  rapidities $\eta$ into the target wave function in principle affects the evolution of the target wave function between rapidity $\eta$ and rapidity $Y$, at which the target scatters on the projectile. However, since in the dense regime the rate of the evolution does not depend on the number of dipoles, there is in fact almost no dependence on $\eta$ as long as at that $\eta$ the target is dense ($P$ is close to unity). Thus $\bar P$ is a nontrivial function of $\eta$ only in the rapidity interval in which $P$ significantly differs form unity. This is clearly illustrated on \fig{solh0}-d.


\begin{figure}[ht]
\begin{tabular}{ c c c}
\epsfig{file=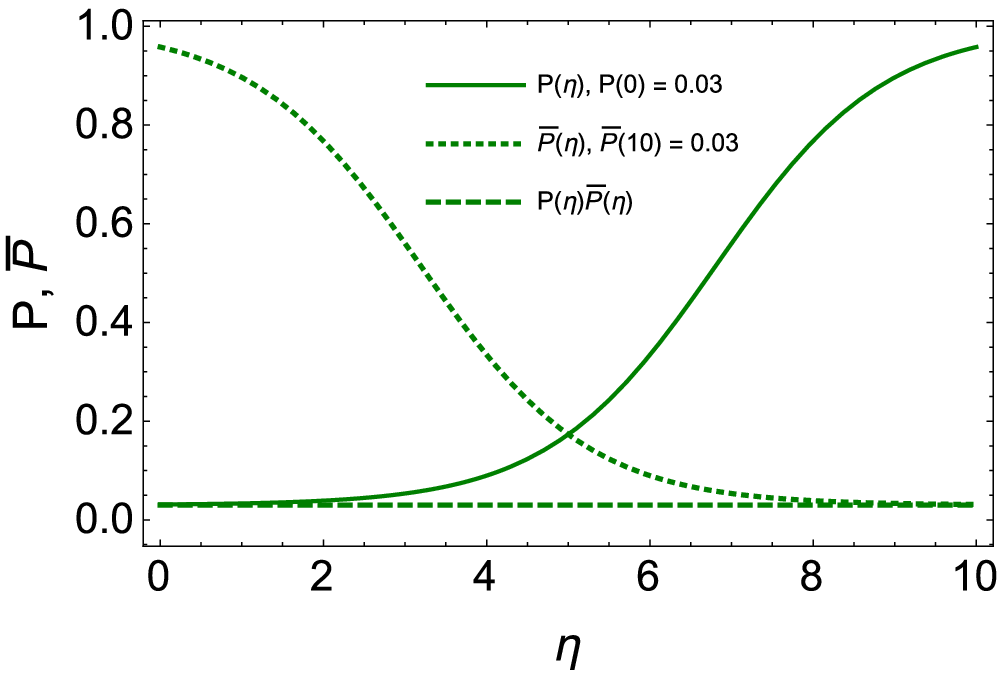,width=75mm} &~~~~~~~&\epsfig{file=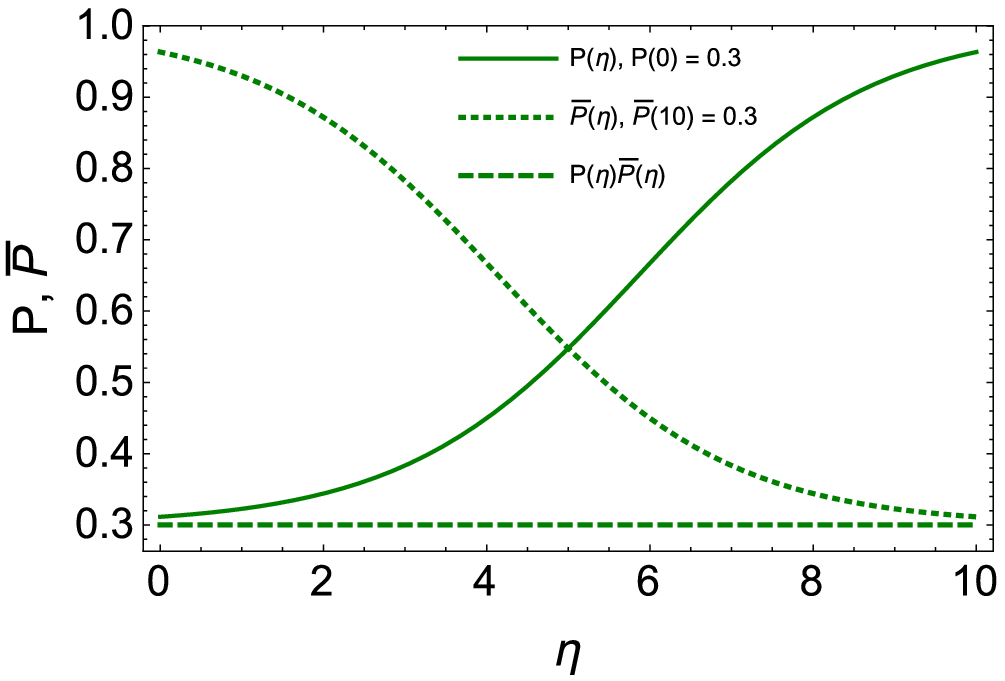,width=75mm} \\
\fig{solh0}-a & & \fig{solh0}-b\\
\epsfig{file=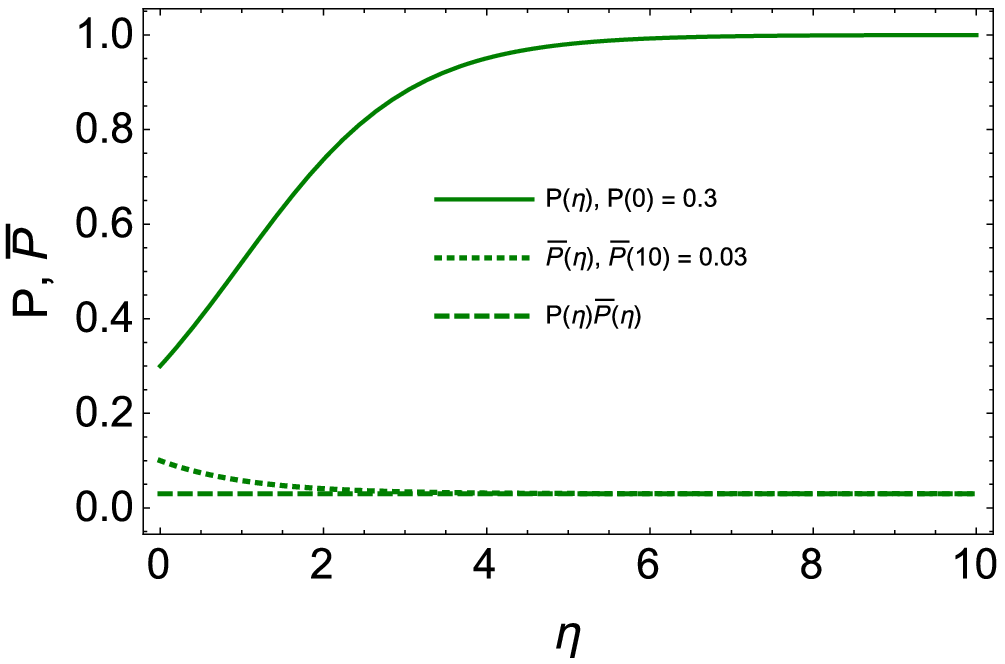,width=75mm} &~~~~~~~&\epsfig{file=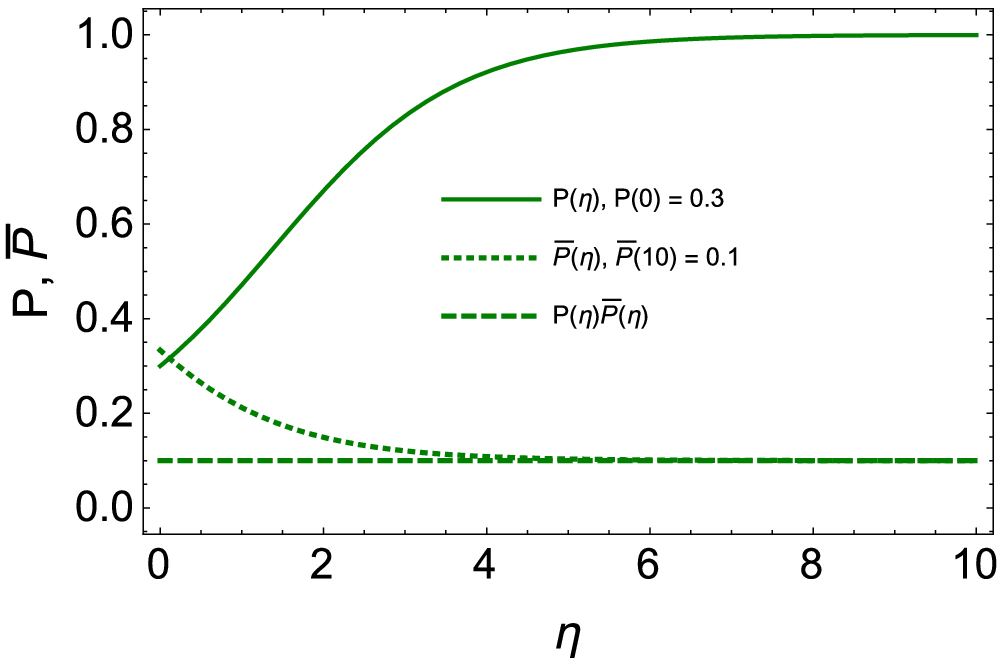,width=75mm} \\
\fig{solh0}-c & & \fig{solh0}-d\\
\epsfig{file=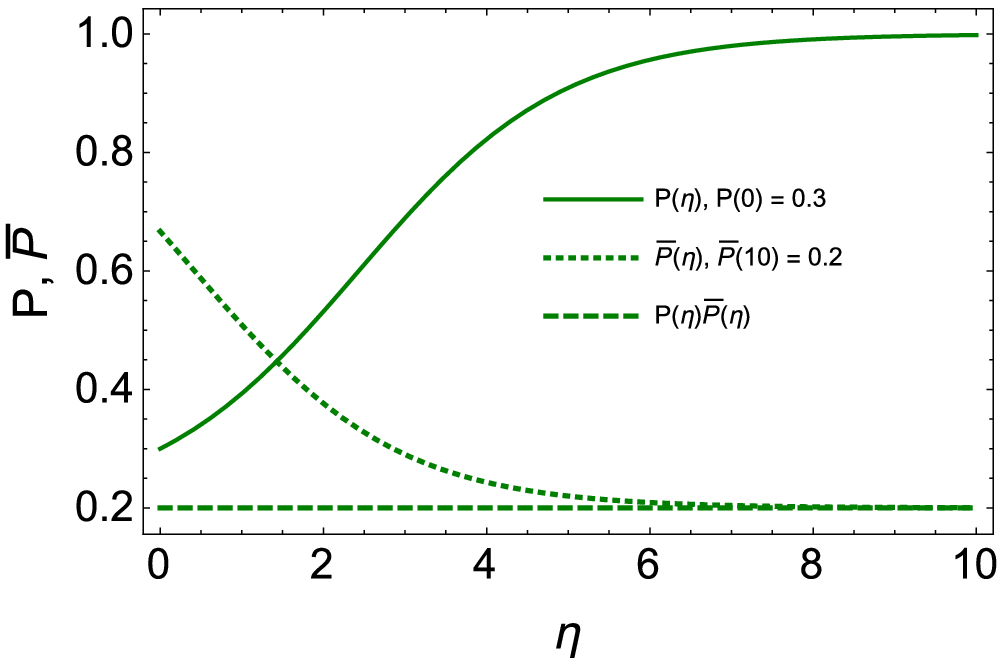,width=75mm} &~~~~~~~&\epsfig{file=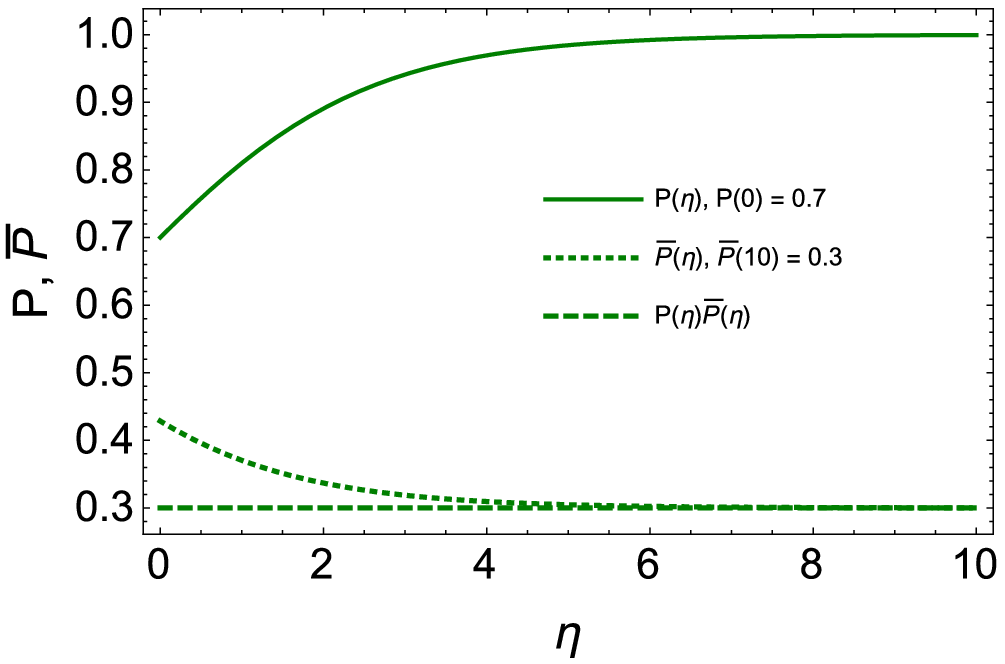,width=75mm} \\
\fig{solh0}-e & & \fig{solh0}-f\\
\end{tabular}
\caption{Examples of numerical solutions of \protect\eq{H01} for Y=10. Considering dipole-dipole amplitude $\gamma\,=\,0.03$  we can assign the following interpretation for these figures.   \protect\fig{solh0}-a: dipole-dipole scattering; \protect\fig{solh0}-b:    scattering of two identical  heavy nuclei; \protect\fig{solh0}-c: dipole-heavy nucleus scattering; \protect\fig{solh0}-d: light nucleus-heavy nucleus scattering; \protect\fig{solh0}-e: scattering of two heavy but different nuclei; and \protect\fig{solh0}-f: very heavy - heavy nucleus scattering.
\label{solh0}}
\end{figure}

 Clearly the classical solutions of eq.(\ref{H01}) determine the scattering amplitude in the semiclassical regime.
To see this explicitly we employ the path integral representation for the $S$-matrix.  The UTM Pomeron Lagrangian is
\beq
L^{UTM}=\int_0^Yd\eta\left[\frac{1}{\gamma} \ln(1-P)\frac{\partial}{\partial \eta}\ln (1-\bar P) -H\right]= \frac{1}{\gamma} \int_0^Yd\eta\left[ \ln(1-P)\frac{\partial}{\partial \eta}\ln (1-\bar P) 
+\bar PP\right]
\eeq
The $S$-matrix is then given by 
\beq
{\cal S}^{UTM}_{m\bar n}(Y)=\int dP(\eta)d\bar P(\eta)e^{\frac{1}{\gamma} \int_0^Yd\eta\left[ \ln(1-P)\frac{\partial}{\partial \eta}\ln (1-\bar P) 
+\bar PP\right]}(1-P(Y))^m(1-\bar P(0))^{\bar n}
\eeq
In the classical approximation\footnote{ In principle, we might be able to also  compute quantum corrections to the classical result}
\begin{eqnarray}\label{classs}
{\cal S}^{UTM}_{m\bar n}(Y)&=&e^{\frac{1}{\gamma} \int_0^Yd\eta\left[ \ln(1-p)\frac{\partial}{\partial \eta}\ln (1-\bar p) 
+\bar pp\right]}[1-p(Y)]^m[1-\bar p(0)]^{\bar n}|_{p(0)=1-e^{-\gamma \bar n};\  \bar p(Y)=1-e^{-\gamma m}}\nonumber\\
&=&[1-p(Y)]^me^{\frac{1}{\gamma}\int_0^Yd\eta \left[\ln(1-\bar p)+\bar p\right]p}
\end{eqnarray}
where $p(\eta)$ and $\bar p(\eta)$ are solutions of the classical equations of motion with the boundary conditions specified in eq.(\ref{classs}).

It is interesting to compare the scattering amplitude given by this expression to that obtained from the BK equation. For the latter we have
\beq
{\cal S}^{BK}_{m\bar n}(Y)=\int dP(\eta)d\bar P(\eta)e^{\frac{1}{\gamma} \int_0^Yd\eta\left[ \ln(1-P)\frac{\partial}{\partial \eta}\ln (1-\bar P) -
\ln (1-\bar P)PP\right]}(1-P(Y))^m(1-\bar P(0))^{\bar n}
\eeq
In the classical approximation
\begin{eqnarray}\label{classs4}
{\cal S}^{BK}_{m\bar n}(Y)&=&e^{\frac{1}{\gamma} \int_0^Yd\eta\left[ \ln(1-p)\frac{\partial}{\partial \eta}\ln (1-\bar p) 
-\ln(1-\bar p)p\right]}[1-p(Y)]^m[1-\bar p(0)]^{\bar n}|_{p(0)=1-e^{-\gamma \bar n};\  \bar p(Y)=1-e^{-\gamma m}}\nonumber\\
&=&[1-p(Y)]^m
\end{eqnarray}
Note that the solution for $\bar P$ is irrelevant for the BK amplitude, which is determined entirely by $P(Y)$. 
 On the other hand the scattering amplitude in UTM does depend on $\bar P$. Nevertheless the two models should be close in 
 the regime where the BK evolution applies.  An example of  numerically computed amplitudes for BK and UTM
 with generic initial conditions is presented on Fig.(\ref{solh1}) for one projectile dipole ($m=1$), $A= 1-{\cal S}_{1\bar n}$.

\begin{figure}[ht]
\begin{tabular}{c c}
\epsfig{file=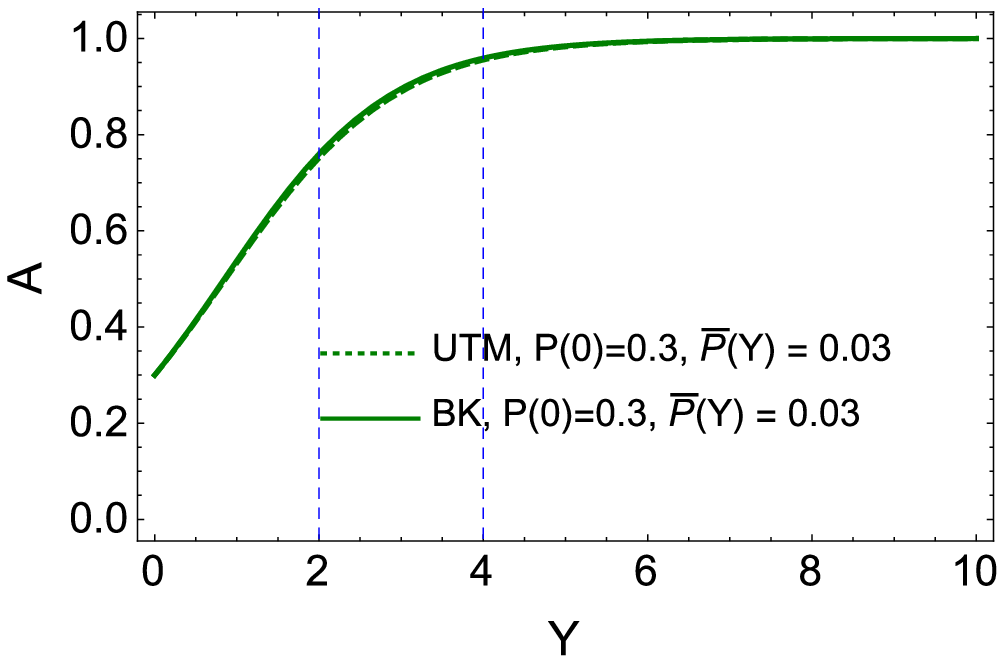,width=80mm} &\epsfig{file=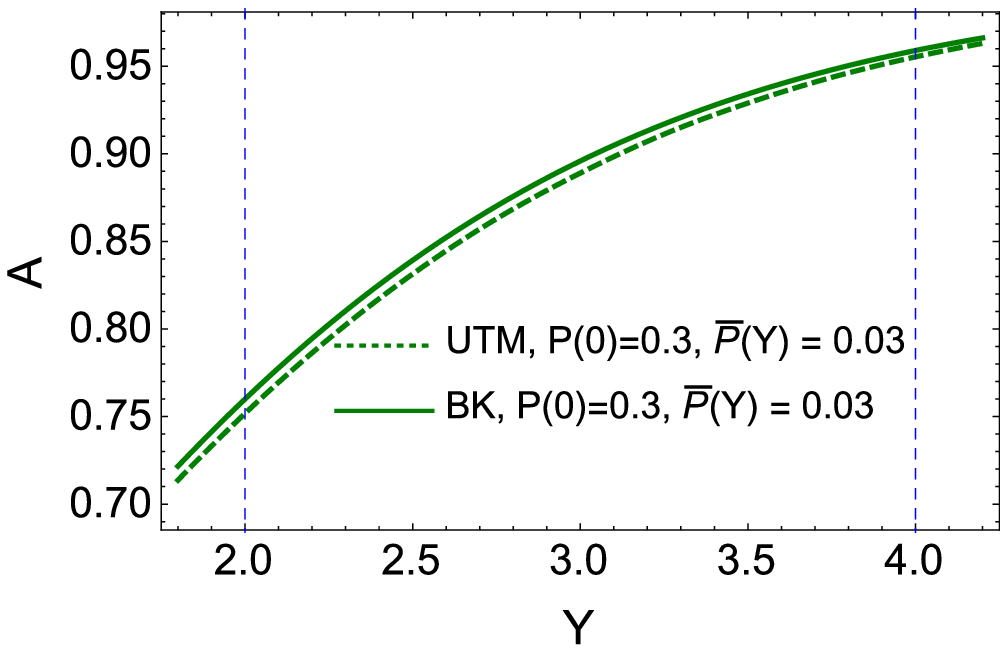,width=80mm}\\
\fig{solh1}-a & \fig{solh1}-b\\
\end{tabular}
\caption{ Scattering amplitude $A=1-{\cal S}_{1\bar n}$ computed for the BK (solid line) and the UTM (dotted line) evolutions for  scattering of a single dipole an a large target (about ten dipoles) with $\gamma=0.03$.   Fig. 2-b zooms into the rapidity interval $2<Y<4$.}
\label{solh1}
\end{figure}

The difference between the two amplitudes is indeed quite small, reaching the maximum of about $\sim 3.5\%$ in the pre-saturation region. 
However, close to the saturation, the difference between the {\it $S$- matrices} of BK and UTM are quite significant, since the $S$-matrix itself is close to zero.
 In other words, unitarization significantly modifies the way the $S$-matrix approaches zero, i.e. the zero dimensional ``Levin-Tuchin law'' \cite{LETU}.

To illustrate this fact we plot the relative difference of the $S$-matrix between the two calculations in Fig.(\ref{smat}).
 We expect that the effect of the unitarization should be even more pronounced in less inclusive observables like particle multiplicity.
\begin{figure}[ht]
\epsfig{file=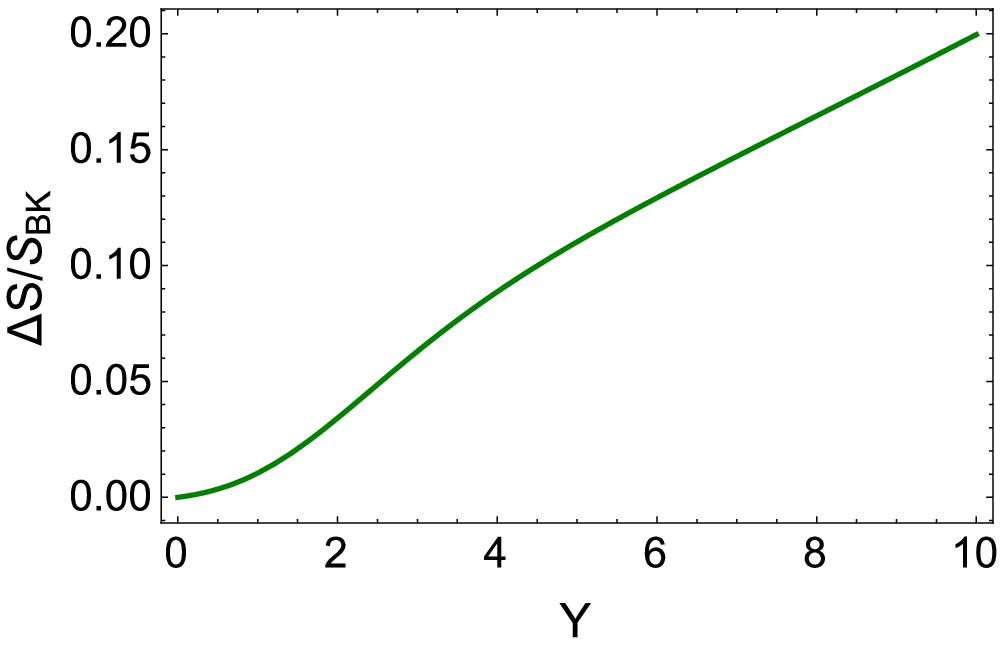,width=100mm} 
\caption{Difference between the scattering matrices for the BK and the UTM evolutions normalized to the BK evolution for 
scattering of a single dipole on a large target with $\gamma=0.03$.   }
\label{smat}
\end{figure}

\section{Two transverse dimensions?}
The next natural step is to try and generalize the considerations of the previous section to the real world of two transverse dimensions.

The logic of changing the commutation relations between $P$ and $\bar P$ is the same as in the toy model.
Thus the correct commutation relations between the Pomeron and its dual are
\beq\label{comrel}
[1-P(x,y)][1-\bar P(u,v)]=(1-\gamma(xy,uv))[1-\bar P(u,v)][1- P(x,y)]
\eeq
In principle, the commutation relations (\ref{comrel}) should be derivable in the large $N_C$ limit from the definitions (\ref{poms}).

In the following we will assume that $P(xy)=P(yx)$, and neglect the possible contribution of Odderon to dipole scattering.
This allows the following representation 
\beq
P(x,y)=1-\exp\{-\Phi(x,y)\}; \ \ \ \ \bar P(x,y)=1-\exp\{\int_{u,v}\gamma(xy;uv)\frac{\delta}{\delta \Phi(u,v)}\}
\eeq
Here like in the zero dimensional case we neglected the difference between $\gamma$ and $\ln (1-\gamma)$, since $\gamma=O(\alpha_s^2)\ll 1$. 
Now recall that the function $\gamma$ satisfies (\ref{gamma2}).
Therefore we can write
\beq 
P^\dagger(x,y)=\frac{32\pi^2}{\alpha_s^2}\frac{1}{d(x,y)}\nabla^2_x\nabla^2_y[\ln\bar d(x,y)]
\eeq

The BK hamiltonian thus can be written as
\beq
H_{BK}=\frac{16\pi N_C}{\alpha_s}\int_{x,y,z}\nabla^2_x\nabla^2_y[\ln \bar d(x,y)]K(x,y,z)\left[1-\frac{d(x,z)d(z,y)}{d(x,y)}\right]
\eeq

This is the representation of the BK Hamiltonian in terms of $d$ and $\bar d$. We stress that this is the original, unaltered BK Hamiltonian but written in terms of $\bar d$ rather than $P^\dagger$.

When written in this form, it may seem that the action of this Hamiltonian is non-unitary both on the projectile and the target, since it contains a logarithm of $\bar d$ as well as the factor $1/d$, and both cannot be expanded in Taylor series. However a little reflection shows that the action on the projectile is in fact unitary. This is because the operator $\nabla^2_x\nabla^2_y[\ln \bar d(x,y)]$ annihilates the projectile wave function unless it contains a dipole with coordinates $(x,y)$. In the latter case this dipole compensates the explicit factor $1/d(x,y)$ and leads to an expression expandable in Taylor series in $d$, since the original wave function itself is expandable. However the action of $H_{BK}$ on the target is indeed non-unitary due to the logarithmic factor just like in the zero dimensional model.

From this point on one would like to modify this Hamiltonian in the same spirit as done in the previous section. The result one aspires to is a Hamiltonian which is 

A. Self dual under transformation $d\rightarrow \bar d$;

B. Unitary - acting on either projectile or target wave function with fixed number of dipoles it creates additional dipoles with positive probability, while the total probability does not change;

C. When expanded in $P^\dagger(x,y)$ to linear order reproduces $H_{BK}$.

 We can satisfy conditions A and C by the following ansatz 
\beq H_B=\frac{16\pi N_C}{\alpha_s}\int_{x,y,z,\bar x,\bar y,\bar z}
K(\bar x,\bar y,\bar z)\left[1-\frac{\bar d(\bar x,\bar z)\bar d(\bar z,\bar y)}{\bar d(\bar x,\bar y)}\right]{\cal L}(\bar x\bar y,x,y)
K(x,y,z)\left[1-\frac{d(x,z)d(z,y)}{d(x,y)}\right]
\eeq
with the operator ${\cal L}$ chosen such that
\beq 
\int_{\bar x,\bar y,\bar z} M(\bar x,\bar y,\bar z)\left[1-\bar d(\bar x,\bar z)-\bar d(\bar z,\bar y)+\bar d(\bar x,\bar y)\right]{\cal L}(\bar x\bar y,xy)=\nabla^2_x\nabla^2_y\bar d(x,y)
\eeq
However we were unable to show that this Hamiltonian generates unitary evolution. The difficulty resides in the non-locality of the relation between $\bar P(x,y)$ and $P^\dagger(x,y)$. A necessary (although not sufficient) condition for unitarity is that after acting on the target wave function, the coordinates $(\bar x,\bar y)$ should become equal to coordinates of one of the dipoles contained in the wave function. If this is the case, then the factor $1/\bar d(\bar x,\bar y)$ cancels, and the result is expandable in Taylor series. However the mutual non-locality of $P^\dagger$ and $\bar P$ makes this very difficult to achieve.
Perhaps some additional physical insight is needed to resolve this question. 

We are therefore forced to postpone this problem until better times.

\section{Conclusions.}
In this paper we have examined the question of {the $s$-channel} unitarity of the QCD Reggeon Field Theory. We have shown, starting from the QCD definition of scattering amplitudes, how the requirement of unitarity of the QCD evolution should be reflected in the action of RFT Hamiltonian on the projectile and target wave functions.

Our finding is that the action of the BK (and JIMWLK) Hamiltonian is unitary on the projectile wave function, but violates unitarity of the target wave function. The Braun Hamiltonian, which is a self dual extension of the BK Hamiltonian, turns out to violate unitarity of both, the projectile and the target. This unitarity violation is small in the regime where the respective Hamiltonians are applicable. However the peculiar behavior of the solutions to Braun Hamiltonian at large rapidities is closely related to this violation of unitarity. 
Recall that classical solutions to the Braun equations of motion bifurcate beyond some critical rapidity $Y_c$ \cite{motyka}. Starting at $Y_c$, the  dependence of either $\bar P$ or $P$ on $\eta$ becomes unphysical, indicating large unitarity violation. Therefore starting from this total rapidity the Braun evolution is not trustworthy as unitarity violating effects are large.

To elucidate the unitarity considerations we have discussed toy models of RFT in zero transverse dimensions. We have found that with correct commutation relations between the Pomeron and its dual, $P$ and $\bar P$ it is possible to modify the Braun evolution to make it unitary. This unitarized toy model (UTM) has many desirable properties. It is self dual, just like the Braun Hamiltonian is, reduces to BK evolution in the limit of dilute projectile and evolves both the projectile and the target wave functions in a unitary way. It also exhibits approach to saturation similar to that we expect in QCD, namely for large dipole density the rate of growth of the dipole number becomes independent of the dipole number itself. 

We found analytic solution of the classical equations of motion of UTM and compared the scattering amplitude calculated in classical approximation to that of the BK model. As expected, the evolution in UTM is somewhat slower, as it takes into account the saturation effects in the projectile wave function. For dilute projectile the difference between the scattering amplitude calculated in BK and UTM models is quite small, 
but the pre-asymptotic behaviour differs significantly in the saturation regime.

In the two dimensional case  we have provided the corrected commutation relations between $P$ and $\bar P$ valid beyond the dilute limit.  Unfortunately so far we were unable to find a modified unitarized version of the Braun Hamiltonian in QCD. This is left for future work.

 Finally we stress that our focus in this paper was on the $s$-channel unitarity. For the RFT to be fully consistent, it has to be $t$-channel unitary as well. The BFKL and Braun Hamiltonians satisfy the $t$-channel unitarity constraints by construction, however their applicability is limited to scattering of dilute systems. On the other hand the BK and JIMWLK Hamiltonians are applicable to scattering processes involving one dilute and one dense system, but they lack the $t$-channel unitarity as discussed in \cite{MShoshi}. It appears that self duality of a Hamiltonian ensures the $t$-channel unitarity of the RFT. Thus in order to generalize the RFT approach to nucleus - nucleus collisions it is imperative to find a  $s$-channel unitary and self dual extension of the BK and Braun Hamiltonians.    

   {\bf Acknowledgements} 
   We are grateful to Al Mueller for discussions and encouragement.  We also thank Physics Departments of
    BGU, UCONN, UTFSM and TAU  for hospitality and our colleagues there for stimulating discussions.
   This research was supported by ISRAELI SCIENCE FOUNDATION grant \#87277111;  the  BSF grant \#2012124; 
 the People Program (Marie Curie Actions) of the European Union's Seventh Framework Program FP7/2007-2013/ under REA
grant agreement \#318921 and  by  the  Fondecyt (Chile) grant  \#1140842.


\begin{thebibliography}{99}


\bibitem{gribov} V.~N.~Gribov,
  Sov.\ Phys.\ JETP {\bf 26}, 414 (1968)
  [Zh.\ Eksp.\ Teor.\ Fiz.\  {\bf 53}, 654 (1967)].


\bibitem{BFKL}
 E. A. Kuraev, L. N. Lipatov, and F. S. Fadin, { Sov. Phys.
JETP}
                {\bf 45}, 199 (1977); \,\,\,
Ya. Ya. Balitsky and L. N. Lipatov,
               { Sov. J. Nucl. Phys.}\, {\bf 28}, 22 (1978).
               
               
\bibitem{glr} L. Gribov, E. Levin and M. Ryskin, Phys. Rept. 100, 1, 1983;

\bibitem{MUPA}
A. H. Mueller and J. Qiu, 
Nucl. Phys. {\bf B268} (1986) 427;\, H. Mueller and B. Patel,
Nucl. Phys. {\bf B425} (1994) 471.

\bibitem{MUDI}
  A.~H.~Mueller,
  Nucl.\ Phys.\ B {\bf 415} (1994) 373;\,\,\,
  Nucl.\ Phys.\ B {\bf 437} (1995) 107;\\
      A.~H.~Mueller and B.~Patel, Nucl. Phys. B 425, 471, 1994.         
               
   \bibitem{LIREV}
                L.~N.~Lipatov,
  Phys.\ Rept.\  {\bf 286} (1997) 131.
               
\bibitem{LipatovFT}
  L.~N.~Lipatov,
  Nucl.\ Phys.\ B {\bf 365}, 614 (1991),
  Nucl.\ Phys.\ B {\bf 452}, 369 (1995), \\
R.~Kirschner, L.~N.~Lipatov and L.~Szymanowski,
  Nucl.\ Phys.\ B {\bf 425}, 579 (1994),
  Phys.\ Rev.\ D {\bf 51}, 838 (1995).


  \bibitem{bartels} J.~Bartels, Z.Phys. {\bf C60}, 471 (1993);\\
J.~Bartels and M.~Wusthoff, Z. Phys. {\bf C66}, 157 (1995);\,\,\,
 J.~Bartels and C.~Ewerz, JHEP  {\bf 9909}, 026 (1999), \\
   C.~Ewerz,
  JHEP {\bf 0104} (2001) 031.

\bibitem{BKP}
J.~Bartels,
%
Nucl.\ Phys.\  {\bf B175}, 365 (1980);\\
J.~Kwiecinski and M.~Praszalowicz,
%
Phys.\ Lett.\  {\bf B94}, 413 (1980).

  
  \bibitem{mv} L.~McLerran and R.~Venugopalan, Phys. Rev. D49: 2233-2241, (1994);  Phys. Rev. D49: 3352-3355, (1994). 

\bibitem{Salam}
A.~H.~Mueller and G.~P.~Salam,
  Nucl.\ Phys.\ B {\bf 475}, 293 (1996);\\
  G.~P.~Salam,
  Nucl.\ Phys.\ B {\bf 461}, 512 (1996).
  
   \bibitem{KOLE}
  Y.~V.~Kovchegov and E.~Levin,
  Nucl.\ Phys.\ B {\bf 577} (2000) 221.
  
   \bibitem{BRN}
M. A. Braun,
Eur. Phys. J. {\bf C16} (2000) 337;\\
M. A. Braun and G. P. Vacca,
Eur. Phys. J. {\bf C6} (1999) 147; \\
J.~Bartels, M.~Braun and G.~P.~Vacca,
  Eur.\ Phys.\ J.\ C {\bf 40}, 419 (2005). 
\\
 J.~Bartels, L.~N.~Lipatov and G.~P.~Vacca,
  Nucl.\ Phys.\ B {\bf 706}, 391 (2005).

\bibitem{braun}  M.~A.~Braun,
  Phys.\ Lett.\ B {\bf 483}, 115 (2000), 
  Eur.\ Phys.\ J.\ C {\bf 33}, 113 (2004);
  Phys.\ Lett.\ B {\bf 632}, 297 (2006).  
  
 
\bibitem{BK}
I.~Balitsky,
{Phys.\ Rev.} {\bf D60}, 014020 (1999);
Y.~V.~Kovchegov,
{Phys.\ Rev.}  {\bf D60}, 034008  (1999).
 
  
   \bibitem{reggeon} A. Kovner and M. Lublinsky, JHEP 0702:058 (2007),
\bibitem{jimwlk} J. Jalilian Marian, A. Kovner, A. Leonidov and H. Weigert,
{Nucl. Phys.}{\bf  B504} 415 (1997);  
{ Phys. Rev.} {\bf D59} 014014 (1999);  \\
J. Jalilian Marian, A. Kovner and H. Weigert, {\it Phys. Rev.}{\bf D59} 
014015 (1999); \\
A. Kovner and J.G. Milhano, {Phys. Rev.} {\bf D61} 014012 (2000); \\
 A. Kovner, J.G. Milhano and H. Weigert,
{ Phys. Rev.} {\bf D62} 114005 (2000); \\
 H. Weigert, { Nucl.Phys.} {\bf A 703} (2002) 823.
 
 \bibitem{cgc}  E.Iancu, A. Leonidov and L. McLerran, {Nucl. Phys.} 
{\bf A 692} (2001) 583; {Phys. Lett.} {\bf B
510} (2001) 133;\\
E. Ferreiro, E. Iancu, A. Leonidov, L. McLerran;  
{Nucl. Phys.}{\bf A703} (2002) 489.
  

\bibitem{ourbraun} T. Altinoluk, A. Kovner, E. Levin and M. Lublinsky, JHEP 1404 (2014) 075.

\bibitem{klwmij} A. Kovner and M. Lublinsky; Phys. Rev. {\bf D71}: 085004 (2005).

\bibitem{KLduality} 
  A.~Kovner and M.~Lublinsky,
  Phys.\ Rev.\ Lett.\  {\bf 94}, 181603 (2005).


\bibitem{Balitsky05}
  I.~Balitsky,
  Phys.\ Rev.\ D {\bf 72}, 074027 (2005).

\bibitem{SMITH}
  Y.~Hatta, E.~Iancu, L.~McLerran, A.~Stasto and D.~N.~Triantafyllopoulos,
  Nucl.\ Phys.\ A {\bf 764}, 423 (2006).
  
\bibitem{foam} A. Kovner, M. Lublinsky and U. Wiedemann, JHEP {\bf 0706}, 075 (2007).
T.~Altinoluk, A.~Kovner, M.~Lublinsky and J.~Peressutti,
  JHEP {\bf 0903}, 109 (2009).
  
\bibitem{GLV} 
  F.~Gelis, T.~Lappi and R.~Venugopalan,
  Phys.\ Rev.\ D {\bf 78}, 054019 (2008).

\bibitem{gluon}
  L.~N.~Lipatov,
  Sov.\ J.\ Nucl.\ Phys.\  {\bf 23}, 338 (1976)
  [Yad.\ Fiz.\  {\bf 23}, 642 (1976)];\\
 L.~L.~Frankfurt and V.~E.~Sherman,
  Sov.\ J.\ Nucl.\ Phys.\  {\bf 23} (1976) 581;\\
V.~S.~Fadin, M.~I.~Kotsky and R.~Fiore,
  Phys.\ Lett.\ B {\bf 359} (1995) 181.


\bibitem{motyka} S. Bondarenko and L. Motyka,  Phys. Rev. D75 (2007) 114015.


\bibitem{MShoshi} 
  A.~H.~Mueller and A.~I.~Shoshi,
  Nucl.\ Phys.\ B {\bf 692}, 175 (2004).

\bibitem{IAN}
  E.~Iancu and D.~N.~Triantafyllopoulos,
  Nucl.\ Phys.\ A {\bf 756} (2005) 419.
  Phys.\ Lett.\ B {\bf 610} (2005) 253; \\
   E.~Iancu, G.~Soyez and D.~N.~Triantafyllopoulos,
  Nucl.\ Phys.\ A {\bf 768} (2006) 194.
  
  \bibitem{MUSH}
    A.~H.~Mueller, A.~I.~Shoshi and S.~M.~H.~Wong,
  Nucl.\ Phys.\ B {\bf 715} (2005) 440.
  
   \bibitem{LELU}
   E.~Levin and M.~Lublinsky,
  Phys.\ Lett.\ B {\bf 607} (2005) 131,
  Nucl.\ Phys.\ A {\bf 763} (2005) 172.

  
  \bibitem{KLP}
    A.~Kormilitzin, E.~Levin and A.~Prygarin,
  Nucl.\ Phys.\ A {\bf 813} (2008) 1.
    \bibitem{LMP}
    E.~Levin, J.~Miller and A.~Prygarin,
  Nucl.\ Phys.\ A {\bf 806} (2008) 245.
  \bibitem{LEPP}
  E.~Levin,
  JHEP {\bf 1311} (2013) 039.



  \bibitem{ACJ}
  D.~Amati, L.~Caneschi and R.~Jengo,
  Nucl.\ Phys.\ B {\bf 101} (1975) 397.
  \bibitem{AAJ}
  V.~Alessandrini, D.~Amati and R.~Jengo,
  Nucl.\ Phys.\ B {\bf 108} (1976) 425.

  \bibitem{JEN}
   R.~Jengo,
  Nucl.\ Phys.\ B {\bf 108} (1976) 447.
    \bibitem{ABMC}
   D.~Amati, M.~Le Bellac, G.~Marchesini and M.~Ciafaloni,
  Nucl.\ Phys.\ B {\bf 112} (1976) 107.
\bibitem{CLR}
 M.~Ciafaloni, M.~Le Bellac and G.~C.~Rossi,
  Nucl.\ Phys.\ B {\bf 130} (1977) 388.
  
  
  \bibitem{CIAF}
    M.~Ciafaloni,
  Nucl.\ Phys.\ B {\bf 146} (1978) 427.
  
 
  \bibitem{RS}
  P.~Rembiesa and A.~M.~Stasto,
  Nucl.\ Phys.\ B {\bf 725} (2005) 251.
  
  \bibitem{KLremark2} 
  A.~Kovner and M.~Lublinsky,
  Nucl.\ Phys.\ A {\bf 767} 171 (2006).

 
  \bibitem{SHXI}  
    A.~I.~Shoshi and B.~W.~Xiao,
  Phys.\ Rev.\ D {\bf 73} (2006) 094014.
  \bibitem{KOLEV}
   M.~Kozlov and E.~Levin,
  Nucl.\ Phys.\ A {\bf 779} (2006) 142.
  \bibitem{BIT}
   J.-P.~Blaizot, E.~Iancu and D.~N.~Triantafyllopoulos,
  Nucl.\ Phys.\ A {\bf 784} (2007) 227.
  .	
\bibitem{nestor} N. Armesto, S. Bondarenko, J. G. Milhano and P. Quiroga, JHEP 0805 (2008) 103.
  
\bibitem{LEPRI}
  E.~Levin and A.~Prygarin,
  Eur.\ Phys.\ J.\ C {\bf 53} (2008) 385.


 \bibitem{yinyang} A. Kovner and M. Lublinsky,  Phys.Rev. D72 (2005) 074023;   Nucl. Phys. A 779:220-243, (2006).
 \bibitem{LETU}
  E.~Levin and K.~Tuchin,
  Nucl.\ Phys.\ B {\bf 573} (2000) 833.
  
  
  \bibitem{brauntarasov}  	
M.A. Braun and  A. Tarasov; Nucl. Phys. B851 (2011) 533-550;
 Nucl. Phys. B863 (2012) 495-509.


  
  \bibitem{shoshi} S. Bondarenko, L. Motyka, A.H. Mueller, A.I. Shoshi and B.-W. Xiao; Eur. Phys.J. C50 (2007) 593-601.





\end{thebibliography}
\end{document}